\documentclass[lettersize,journal]{IEEEtran}
\usepackage{amsmath,amsfonts}
\usepackage{booktabs}
\usepackage{multirow}
\usepackage{algorithmic}
\usepackage{algorithm}
\usepackage{array}
\usepackage[caption=false,font=normalsize,labelfont=sf,textfont=sf]{subfig}
\usepackage{textcomp}
\usepackage{stfloats}
\usepackage[hyphens]{url}
\usepackage{verbatim}
\usepackage{graphicx}
\usepackage{graphics}
\usepackage{float}
\usepackage{makecell}
\usepackage{cite}
\usepackage{etoolbox}
\usepackage{nomencl}
\makenomenclature
\renewcommand{\nomgroup}[1]{%
  \item[\bfseries
  \ifstrequal{#1}{A}{Sets and Indices}{%
  \ifstrequal{#1}{B}{Parameters and Constants}{%
  \ifstrequal{#1}{C}{Variables}{%
  \ifstrequal{#1}{D}{Acronyms}{}%
  }}}]}
\usepackage{enumitem}
\usepackage{comment}
\usepackage{xcolor}
\usepackage{hyperref}
\hypersetup{
colorlinks,
citecolor=blue,
linkcolor=blue,    
urlcolor=blue,
}
\usepackage{tabstackengine}
\usepackage{scalerel}
\usepackage{stackengine}
\usepackage{accents}
\usepackage[english]{babel}
\usepackage{amsthm}

\DeclareMathAlphabet{\mymathbb}{U}{BOONDOX-ds}{m}{n}
\newcommand{\todo}[1]{\textcolor{red}{{\bf Todo:}  #1}}

\usepackage{tabularx}
\usepackage{makecell}
\newcolumntype{Y}{>{\centering\arraybackslash}X}

\begin{document}

\title{
  Power Estimation and Optimal Work--Charging Scheduling of Construction Electric Vehicles via Mobile Charging Stations
{\footnotesize \textsuperscript{}}
}
\author{Avik~Ghosh*, 
        Akın~Taşcıkaraoğlu*,
        Daniela~Rojas, 
        Muhammed~A.~Beyazıt, 
        Mohammad~Reza~Salehizadeh,
        Keaton~Chia,
        Sasha~Doppelt,
        Michael~Ferry, 
        Jan~Kleissl,
        Sujit~Dey, 
        and Yuanyuan~Shi 
        \vspace{-2em}
\thanks{* A.~Ghosh and A.~Taşcıkaraoğlu contributed equally. \\
This work was supported by the California Energy Commission under the award number EPC-24-031. A. Taşcıkaraoğlu was also supported by the TÜBİTAK 2219 International Postdoctoral Research Fellowship Program and the TÜBA Distinguished Young Scientist Award. M. A. Beyazıt was supported by the Scientific and Technological Research Council of Türkiye (TÜBİTAK) Directorate of Science Fellowships and Grant Programmes (BİDEB) 2211 National PhD Scholarship Program. M. R. Salehizadeh was supported through the QRRRF-funded Driving Resilience project (CQU.0001.2324D.RFI).   
\\ A.~Ghosh, D.~Rojas, S.~Dey and Y.~Shi are with the Department of Electrical and Computer Engineering, University of California, San Diego, CA 92093 USA (UCSD). 
Email:\{avghosh,d2rojas,sdey,yyshi\}@ucsd.edu \\
A.~Taşcıkaraoğlu is with the Department of Energy Systems Engineering, İzmir Institute of Technology, 35430 Urla, İzmir, Türkiye. Email: akintascikaraoglu@iyte.edu.tr  \\
M. A. Beyazıt is with the Department of Electrical and Electronics Engineering, Muğla Sıtkı Koçman University, Muğla, 48000,
Türkiye. Email: muhammedalibeyazit@gmail.com \\
M. R. Salehizadeh is with the School of Engineering and Technology, Central Queensland University, Rockhampton, QLD, Australia. Email: m.salehizadeh@cqu.edu.au \\
K.~Chia, S.~Doppelt, M.~Ferry and J.~Kleissl are with the Center for Energy Research at UCSD.
Email: \{kwchia,sdoppelt,mdferry,jkleissl\}@ucsd.edu}}

\maketitle
\flushbottom%

\begin{abstract}
Construction electric vehicles (CEVs) are a promising clean alternative to diesel-powered construction equipment, but their adoption is constrained by sparse onsite charging infrastructure, limited CEV mobility, and insufficient understanding of their power consumption. We address these gaps through a field-data-driven framework coupling CEV power estimation with mobile-charging-aware work scheduling. First, using a real-world construction demonstration at the University of California, San Diego, we develop and validate a per-subactivity power estimation model for a compact electric excavator. Manually labeled video is synchronized with coarse battery state-of-charge (SOC) telematics, and constrained nonnegative least squares is used to recover each subactivity's average power consumption. The model predicts held-out test data within $17\%$ normalized mean absolute error (NMAE), and the accompanying dataset is released publicly. Second, leveraging the subactivity power estimates, we formulate a mixed-integer program that jointly optimizes CEV work and charging schedules together with the location, timing, and charging/discharging of mobile charging stations (MCSs) serving the CEVs. The optimization accounts for energy and demand charges, carbon emissions, unmet work penalties, MCS travel, and the physical and operational constraints of the CEVs and MCSs. Across realistic scenarios drawn from the demonstration, the proposed co-optimization attains the lowest operating cost in every case, being $7$--$96\%$ below the best-performing baseline, while solving most instances to proven optimality within an hour. Dataset and scripts are available at \url{https://github.com/ghosh-avik/CEV-MCS-Power-Estimation-and-Joint-Scheduling}. 
\end{abstract}


\nomenclature[A01]{$\mathcal{M}$}{Set of mobile charging stations (MCSs)}
\nomenclature[A02]{$\mathcal{E}$}{Set of construction electric vehicles (CEVs)}
\nomenclature[A03]{$\mathcal{N}$}{Set of all nodes, $\mathcal{N}=\mathcal{N}^g \cup \mathcal{N}^c$}
\nomenclature[A04]{$\mathcal{N}^g$}{Set of grid (depot) nodes}
\nomenclature[A05]{$\mathcal{N}^c$}{Set of construction nodes}
\nomenclature[A06]{$\mathcal{A}$}{Set of separately identifiable CEV subactivities}
\nomenclature[A07]{$\mathcal{A}^{\mathrm{con}}$}{Set of productive construction subactivities, $\mathcal{A}^{\mathrm{con}}\subseteq\mathcal{A}$}
\nomenclature[A08]{$\mathcal{T}$}{Set of all time intervals}
\nomenclature[A09]{$\mathcal{T}^+$}{Set of time indices including the initial instant $t=0$, i.e., $\mathcal{T}^{+}=\{0\}\cup\mathcal{T}$}
\nomenclature[A10]{$\mathcal{T}^{\mathrm{op}}$}{Set of on-peak time intervals}
\nomenclature[A11]{$\mathcal{T}^{\mathrm{work}}$}{Set of working time intervals}
\nomenclature[A12]{$\mathcal{D}$}{Set of all decision variables of the optimization: includes $P_{m,i,t}^{\mathrm{ch,MCS}}$, $P_{m,i,t}^{\mathrm{dch,MCS}}$, $P_{m,i,e,t}^{\mathrm{MCS}\rightarrow\mathrm{CEV}}$, $P_{i,e,t}^{\mathrm{work}}$, $s_{i,a}^{\mathrm{miss}}$, $u_{i,e,t,a}$, $\rho_{m,i,e,t}$, $\beta_{m,i,t}^{\mathrm{arr}}$, $\beta_{m,i,t}^{\mathrm{dep}}$, $x_{m,i,j,t}$, $y_{m,i,j,t}$, $\mu_{i,e,t}$, $z_{m,i,t}$} 
\nomenclature[A13]{$m,\,e$}{Indices of MCSs and CEVs}
\nomenclature[A14]{$i,\,j$}{Indices of nodes}
\nomenclature[A15]{$a,\,a'$}{Indices of subactivities}
\nomenclature[A16]{$t$}{Index of time interval}

\nomenclature[B01]{$C_{\mathrm{batt}}$}{Usable CEV battery capacity (kWh)}
\nomenclature[B02]{$p_a$}{Average power drawn during subactivity $a$ (kW)}
\nomenclature[B05]{$\tau$}{Cumulative SOC-drop threshold for observation window (\%)}
\nomenclature[B06]{$d$}{Number of energy-balance observation windows}
\nomenclature[B07]{$K,\,R$}{Number of cross-validation folds and repetitions}
\nomenclature[B08]{$\Delta T$}{Scheduling time-step duration (h)}
\nomenclature[B09]{$\lambda_t^{\mathrm{elec}}$}{Electricity charge rate over time interval $t$ (\$/kWh)}
\nomenclature[B10]{$\lambda_t^{\mathrm{CO_2}}$}{Grid carbon-emissions intensity over time interval $t$ (kg\,CO$_2$/kWh)}
\nomenclature[B11]{$\lambda^{\mathrm{em}}$}{Carbon-emissions-to-dollar conversion factor (\$/kg\,CO$_2$)}
\nomenclature[B12]{$\lambda^{\mathrm{NC}}$}{Non-coincident demand-charge rate (\$/kW)}
\nomenclature[B13]{$\lambda^{\mathrm{OP}}$}{On-peak demand-charge rate (\$/kW)}
\nomenclature[B14]{$\rho^{\mathrm{miss}}$}{Missed-work penalty (\$/h)}
\nomenclature[B15]{$\rho^{\mathrm{travel}}$}{MCS towing labor cost (\$/h)}
\nomenclature[B16]{${\text {CH}}_m^{\mathrm{MCS}}$}{Grid-charging power capacity of MCS $m$ (kW)}
\nomenclature[B17]{${\text {DCH}}_m^{\mathrm{MCS}}$}{Discharging power capacity of MCS $m$ (kW)}
\nomenclature[B18]{${\text{DCH}}_m^{\mathrm{plug}}$}{Per-plug discharging power limit of MCS $m$ (kW)}
\nomenclature[B19]{$C_m^{\mathrm{plug}}$}{Number of outlet plugs of MCS $m$}
\nomenclature[B20]{${\text{CH}}_e^{\mathrm{CEV}}$}{Charging acceptance rate of CEV $e$ (kW)}
\nomenclature[B21]{$\eta_m/\eta_e$}{Charging/discharging efficiency of MCS $m$ / CEV $e$}
\nomenclature[B23]{$A_{i,e}$}{CEV-to-node assignment indicator}
\nomenclature[B24]{$R_{e,t}$}{Work-power capacity of CEV $e$ over time interval $t$ (kW)}
\nomenclature[B25]{$H_{i,a}$}{Required productive work duration for subactivity $a$ at node $i$ (h)}
\nomenclature[B26]{$\tau_{i,j}^{\mathrm{trv}}$}{Travel time from node $i$ to node $j$ (intervals)}
\nomenclature[B27]{$\gamma_{a\rightarrow a'}$}{Precedence-relation indicator between subactivities $a$ and $a'$}
 \nomenclature[B28]{$\kappa^{\mathrm{seq}}_{a\rightarrow a'}$}{Subactivity precedence ratio between subactivities $a$ and $a'$}
 \nomenclature[B29]{$\kappa^{\mathrm{wt}}$}{Productive-work-to-travel interval ratio}
 \nomenclature[B30]{$t_{\mathrm{limit}}$}{CEV work limit before mandatory rest (intervals)}
 \nomenclature[B31]{$a^{\mathrm{trv}}$}{Traveling subactivity}

\nomenclature[C02]{$D$}{Subactivity-duration design matrix}
\nomenclature[C03]{$p,\,\hat{p}$}{Subactivity-power vector and its estimate (kW)}
\nomenclature[C04]{$b,\,b_i$}{Window-energy vector and energy of window $i$ (kWh)}
\nomenclature[C05]{$\epsilon,\,\epsilon_i$}{Residual (modeling-error) vector and its $i$th entry}
\nomenclature[C07]{$\mathrm{SOE}_{m,t}^{\mathrm{MCS}}$}{State of energy of MCS $m$ at time $t$ (kWh)}
\nomenclature[C08]{$\mathrm{SOE}_{e,t}^{\mathrm{CEV}}$}{State of energy of CEV $e$ at time $t$ (kWh)}
\nomenclature[C09]{$P_{m,t}^{\mathrm{ch,tot}}$}{Total grid-charging power of MCS $m$ over time interval $t$ (kW)}
\nomenclature[C10]{$P_{m,t}^{\mathrm{dch,tot}}$}{Total discharging power of MCS $m$ over time interval $t$ (kW)}
\nomenclature[C11]{$P_{m,i,t}^{\mathrm{ch,MCS}}$}{Charging power of MCS $m$ at grid node $i$ over time interval $t$ (kW)}
\nomenclature[C12]{$P_{m,i,t}^{\mathrm{dch,MCS}}$}{Discharging power of MCS $m$ at construction node $i$ over time interval $t$ (kW)}
\nomenclature[C13]{$P_{m,i,e,t}^{\mathrm{MCS}\rightarrow\mathrm{CEV}}$}{Power transferred from MCS $m$ to CEV $e$ at node $i$ over time interval $t$ (kW)}
\nomenclature[C14]{$P_{i,e,t}^{\mathrm{work}}$}{Work (construction) power of CEV $e$ over time interval $t$ (kW)}
\nomenclature[C15]{$P^{\mathrm{NC}}$}{Non-coincident demand peak (kW)}
\nomenclature[C16]{$P^{\mathrm{OP}}$}{On-peak demand peak (kW)}
\nomenclature[C19]{$s_{i,a}^{\mathrm{miss}}$}{Missed (unmet) work duration for subactivity $a$ at node $i$ (h)}
\nomenclature[C20]{$u_{i,e,t,a}$}{Binary: 1 iff CEV $e$ performs subactivity $a$ at node $i$ over time interval $t$}
\nomenclature[C21]{$\mu_{i,e,t}$}{Binary: 1 iff CEV $e$ is ready to accept charge at node $i$ over time interval $t$}
\nomenclature[C22]{$\rho_{m,i,e,t}$}{Binary: 1 iff MCS $m$ is connected to CEV $e$ at node $i$ over time interval $t$}
\nomenclature[C23]{$x_{m,i,j,t}$}{Binary: 1 iff MCS $m$ departs node $i$ toward node $j$ over time interval $t$}
\nomenclature[C24]{$y_{m,i,j,t}$}{Binary: 1 iff MCS $m$ is in transit on path $(i,j)$ over time interval $t$}
\nomenclature[C25]{$z_{m,i,t}$}{Binary: 1 iff MCS $m$ is present at node $i$ over time interval $t$}
\nomenclature[C27]{$\beta_{m,i,t}^{\mathrm{arr}}$}{Binary: 1 iff MCS $m$ arrives at node $i$ over time interval $t$}
\nomenclature[C28]{$\beta_{m,i,t}^{\mathrm{dep}}$}{Binary: 1 iff MCS $m$ departs node $i$ over time interval $t$}



\section{Introduction}\label{sec:intro}
\subsection{Motivation}\label{subsec:motivation}
Diesel-powered construction equipment contributes about $1.1\%$ of global $\mathrm{CO_2}$ emissions~\cite{idtech_1}, besides $\mathrm{NO_x}$ and particulate-matter emissions. The construction sector in the UK reports the highest number of occupational cancer cases of any industry, with roughly $8\%$ attributable to diesel-exhaust exposure~\cite{idtech_2}. Tightening U.S. and EU emissions standards and associated incentive programs~\cite{core} are thus accelerating the transition to construction electric vehicles (CEVs).

This transition is constrained by limited charging solutions and insufficient understanding of CEV power consumption. CEVs can demand very high charging power which can impose large peak loads 
on primary substations, and cause overloading~\cite{jansson2025charging}. Accommodating CEV charging would therefore necessitate costly distribution grid upgrades. Moreover, the poor on-road mobility of construction vehicles makes reliance on fixed charging infrastructure impractical~\cite{immobile}. Mobile charging stations (MCSs), i.e., large batteries towed between grid-connection points and construction sites, are an attractive alternative that decouples CEV charging from fixed infrastructure and can shift grid power draw toward low-cost, low-congestion periods. Exploiting this flexibility, however, requires coordinating \emph{when and how much} each CEV works and charges, which hinges on knowing the power consumption of the CEV while working, with \emph{when, where, and by how much} the MCS charges from the grid. These are the two problems which we address in this work.

\subsection{Literature Review}\label{subsec:lit_review} 

Unlike standard EVs, off-road CEVs have received relatively little attention in the smart grid research literature. A main reason is that CEV power consumption is dictated by the \emph{work duty cycle}~\cite{wallander2023electric}, which comprises distinct subactivities and varies widely across CEV types~\cite{lajunen2016nrmm}. Prior studies characterize the energy use and duty cycles of construction equipment largely for \emph{diesel} powertrains---to guide fuel-efficiency and hybridization design~\cite{excavatorFuel2017}, build activity-based fuel and emission inventories from engine maps and real-world telematics~\cite{realworldEF2021,earthworkEnergy2022}, or predict aggregate machine energy and emissions from operating features such as load factor and cycle time~\cite{excavatorCO2NN2017}. 
Among CEVs, existing studies characterize the energy and operating behavior of battery electric excavators for powertrain design~\cite{hao2023energy}. Grid oriented studies quantify the charging requirements and distribution grid impact of electric heavy construction equipment~\cite{jansson2025charging}. None of these, however, identifies the per-subactivity \emph{electrical} power demand of a CEV, which is essential for accurate real-time charging and work scheduling. Such a granular, subactivity-resolved power consumption model is neither published by manufacturers nor available in the literature. 

On the charging side, mobile charging and energy transfer solutions have been proposed to relieve fixed charging infrastructure limitations~\cite{wang2016spatio}. A rich smart grid literature optimizes the routing and scheduling of such mobile assets in distribution systems: mobile power sources are routed and scheduled for resilience enhancement~\cite{lei2019routing}, mobile-storage fleets are dispatched over rolling horizons for service restoration~\cite{yao2020rolling}, and MCSs are coordinated with de-icing robots, repair crews, and emergency generators in multi-stage post-disaster restoration~\cite{an2026time}. MCS scheduling has been extended to coupled distribution--transportation networks under uncertainty~\cite{liu2021stochastic}, storage-transportation scheduling in power systems~\cite{sun2016lagrangian}, multistage robust routing~\cite{lu2022multistage}, and power exchange with EVs and distribution network with a focus on increasing MCS profitability~\cite{mcsRouting2025}. Multi-robot vehicle-to-vehicle (V2V)/vehicle-to-grid (V2G) charging networks have also been planned and coordinated for distribution-system voltage regulation~\cite{sihai2026planning}. Closer to our setting, mobile charging infrastructure has been jointly routed and scheduled for vehicle-to-vehicle energy transfer~\cite{kabir2021joint}, coordinated to provide EV charging service alongside energy arbitrage under time-varying prices~\cite{he2026coordinated}, and assigned to EV charging tasks through auction-driven deep reinforcement learning~\cite{do2026sustainable}. 
In all of these studies, however, the demand served by the mobile charging resources is \emph{exogenous}: critical loads, EV charging requests, and associated service requirements are provided as inputs. For CEV charging, the charging demand is difficult to specify a priori. The problem is further complicated when multiple CEVs request service from the same MCS.

Conversely, the construction operations literature optimizes the \emph{work} of such equipment: electric construction machinery fleets are configured and scheduled for energy efficient task allocation with stationary charging~\cite{huang2024energy}, and portable energy supplies have been proposed for electric off-road machinery~\cite{wallander2023electric}, but neither optimizes the grid-side location, timing, and charging/discharging of the MCS. The CEV work schedule and the MCS charging/path-planning problem have thus been studied only in isolation. To the best of our knowledge, no prior work jointly optimizes \emph{when and which subactivities a CEV performs} together with \emph{where, when, and how much its MCS charges from the grid and discharges to the CEVs}. This paper closes that gap: building on the identified subactivity powers, we co-optimize CEV work scheduling with MCS charging under combined economic and environmental objectives, enabling the efficient opportunity charging of CEVs and MCSs.

\subsection{Contributions}\label{subsec:contributions}
The contributions of the present work are as follows:
\begin{enumerate}
  \item Using a real-life construction demonstration at UC San Diego, we identify the distinct subactivities of a CEV and build and validate a per-subactivity power estimation model across three working materials---soil, decomposed granite, and sand. We also release the dataset associated with the CEV power estimation model publicly~\cite{cev_mcs_github}. To the best of the authors' knowledge, this is the first public dataset and accompanying per-subactivity power estimation analysis for CEVs. 

  \item Based on the subactivity power estimates, we jointly optimize CEV work and charging schedules with the optimal location, timing, routing, and charging of MCSs under economic and environmental objectives. To the best of the authors' knowledge, this is the first framework to jointly optimize CEV work and charging schedules together with MCS routing and charging.
  
  \item We demonstrate the superior cost-saving potential of our proposed MCS-CEV joint optimization strategy against other baseline methods through realistic simulations. {The proposed co-optimization attains the lowest operating cost in every case, being $7$--$96\%$ below the best-performing baseline}. 
  \end{enumerate}

The rest of the paper is organized as follows.
Section~\ref{sec:power_consumption} presents the CEV subactivity identification, power estimation model, and validation. Section~\ref{sec:problem} formulates the joint optimization of CEV work and charging schedules with MCS charging/discharging. Section~\ref{sec:case_study} describes the case study setup and the baselines we compare our optimal strategy with, while Section~\ref{sec:results} presents the results and discussion. Section~\ref{sec:conclusions} concludes the study, highlighting its main takeaways and directions for future work. 

\section{CEV power consumption model}
\label{sec:power_consumption}

\begin{figure*}[t]
\centering
\includegraphics[width=0.7\textwidth]{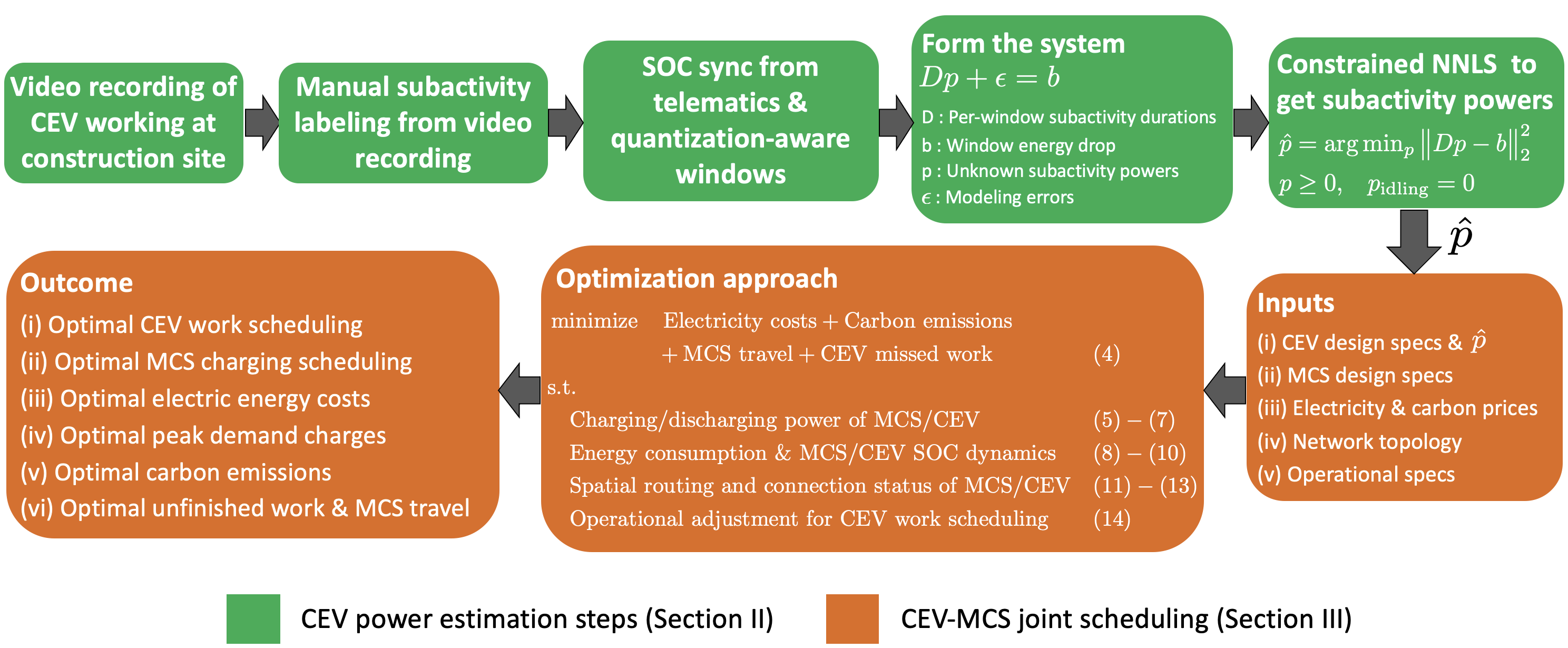}
\vspace{-1 em}
\caption{End-to-end pipeline of the proposed framework. 
}
\vspace{-1.5 em}
\label{fig:pipeline}
\end{figure*}

In this section, we present the methodology and validation analysis for the subactivity power estimation model of the CEV, the various steps of which are demonstrated in Fig.~\ref{fig:pipeline}.

\begin{figure}[!h]
\centering
\subfloat[]{\includegraphics[width=0.4\columnwidth]{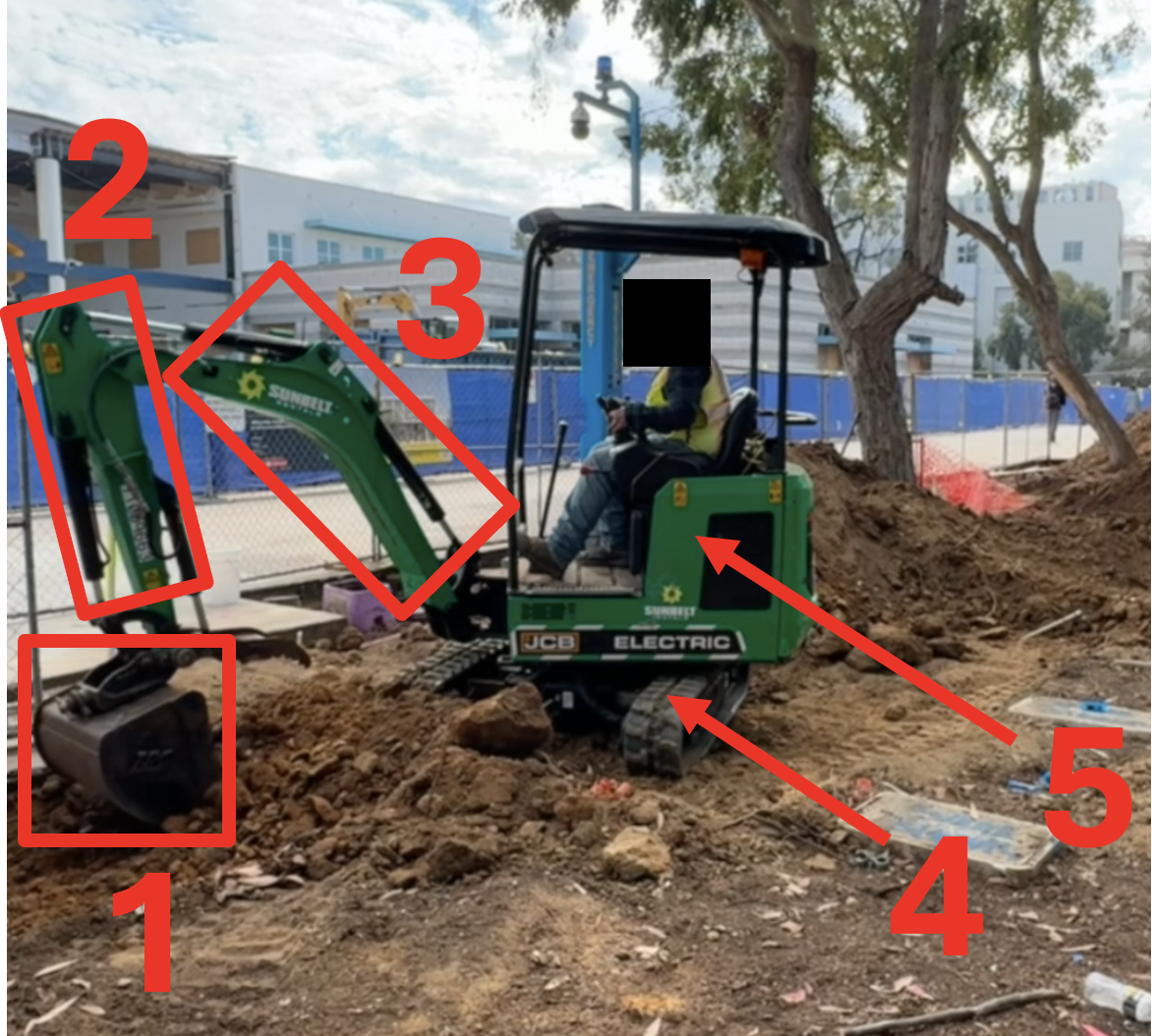}%
\label{fig:CEV-a}}
\hfil
\subfloat[]{\includegraphics[width=0.4\columnwidth]{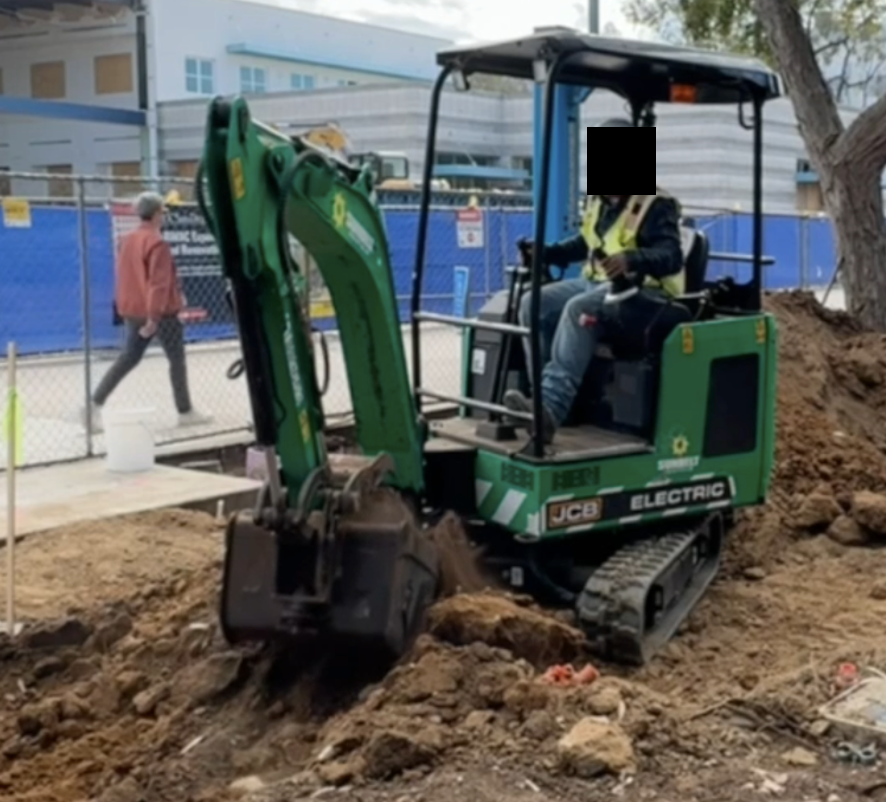}%
\label{fig:CEV-b}}
\\
\subfloat[]{\includegraphics[width=0.4\columnwidth]{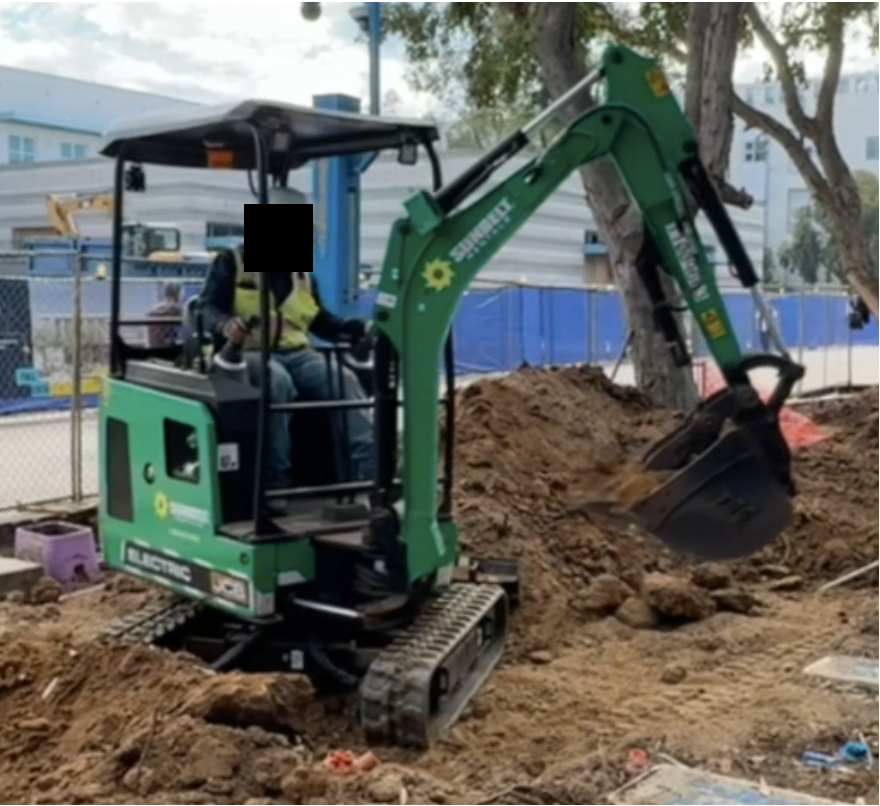}%
\label{fig:CEV-c}}
\hfil
\subfloat[]{\includegraphics[width=0.4\columnwidth]{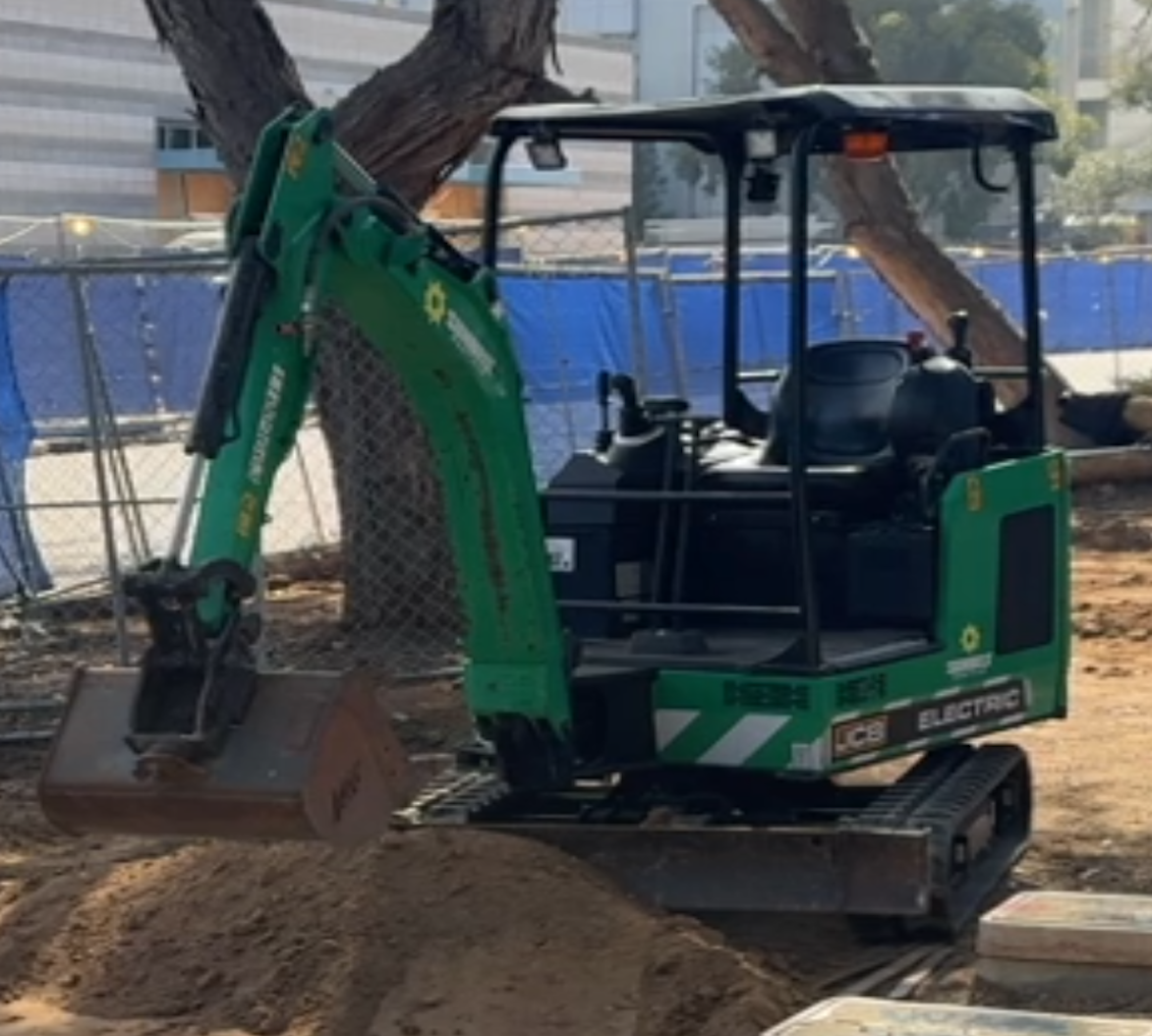}%
\label{fig:CEV-d}}
\caption{A JCB 19C-1E electric compact excavator performs (a) digging; (b) loading; (c) swinging; and (d) idling. The various parts of the CEV in (a) are, 1: Bucket, 2: Arm, 3: Boom, 4: Traveling belt, 
and 5: Swinging body.}
\vspace{-0.1 em}
\label{fig:CEV_subactivity}
\end{figure} 

\subsection{Field Data Collection}
\label{subsec:sites_description}
The studied CEV is a JCB 19C-1E electric compact excavator~\cite{jcb2019quickstart} with a usable battery capacity of $C_{\mathrm{batt}}=14.8$~kWh. {While this excavator is used as the main experimental platform, the proposed methodology is general and can be readily applied to other types of CEVs.} Its on-board telematics record the battery
state-of-charge (SOC) at integer-percent resolution and with non-uniform temporal
granularity, ranging from
seconds to several hours.
Data (i.e., video recording of CEVs working and SOC telematics) were collected during two field campaigns on the UC San Diego campus in October~2025 and February~2026. The campaigns spanned $\approx20$ hours in which the excavator handled three working materials exclusively: \emph{soil}, \emph{decomposed granite}, and \emph{sand}.

\subsection{Subactivity Labeling and SOC Synchronization}
\label{subsec:video_labeling}
Each operating session was recorded on video, as illustrated in Fig.~\ref{fig:CEV_subactivity}. The excavator subactivities were recognized based on the positions and movements of its bucket, arm, boom, swinging body, and traveling belts; see Fig.~\ref{fig:CEV_subactivity} and~\cite[Fig.~1]{molaei2024automatic}. Consistent with the literature~\cite{molaei2024automatic}, we identified five major subactivities: digging, loading, swinging, traveling, and idling. We then manually reviewed the video recording of the construction activities and divided each operating session into contiguous subactivity segments, recording the subactivity class and the actual start and end timestamps of each segment, and hence its duration, at one-second resolution. Each segment was also aligned with the concurrently observed battery SOC, which was held constant between consecutive SOC observations. The resulting manually labeled subactivity taxonomy is described in Table~\ref{tab:subactivity_taxonomy} in Appendix~\ref{appendix_subactivity_classification}.

\subsection{Subactivity Power Estimation Model}
\label{subsec:power_estimation_model}

In this section, we present the power estimation model of the different subactivities of the CEV. 

\subsubsection{Energy-Balance Formulation}\label{subsubsec:energy_balance}

Partition an operating session into $d$ observation windows, indexed by $i$. {An observation window is a period of operation between two battery SOC readings, over which the measured energy drop is attributed to the subactivities performed within it.} During observation window $i$, the battery SOC drops by $\Delta\mathrm{SOC}_i$, delivering energy $b_i=-(\Delta\mathrm{SOC}_i/100)\,C_{\mathrm{batt}}$. Assuming each subactivity $a$ in the set $\mathcal{A}$ of subactivities draws a constant average power $p_a$ (kW) for a duration $D_{i,a}$ (hours), this energy balances as $b_i=\sum_{a\in\mathcal{A}} p_a D_{i,a}+\epsilon_i$ with additive residual $\epsilon_i$. Stacking all $d$ windows yields the linear system
\begin{equation}
D\,p + \epsilon = b, \qquad D\in\mathbb{R}^{d\times |\mathcal{A}|},\;
p\in\mathbb{R}^{|\mathcal{A}|},\; b\in\mathbb{R}^{d},
\label{eq:system}
\end{equation}
where the rows of the design matrix $D$ hold the per-window subactivity durations, the vector $p$ holds the unknown per-subactivity powers to be identified, $b$ collects the measured energies, and $\epsilon$ is the modeling residual.
Equation~\eqref{eq:system} is written with an
explicit residual because the system is over-determined and noisy: with
$d \gg |\mathcal{A}|$, no $p$ satisfies $Dp = b$ exactly.

\subsubsection{Quantization-Aware Observation Window}
\label{subsubsec:window}
Because the SOC telematics are quantized to integer percent, forming one observation window per $1\%$ SOC drop leaves a worst-case rounding error (${\pm}0.5\%$ at each endpoint) as large as the signal itself. {We therefore build the observation windows for~\eqref{eq:system} in a \emph{quantization-aware} manner: we form a new window only once the cumulative SOC has dropped by at least a threshold $\tau$, so that each window's energy drop is large relative to this fixed endpoint error. The next window then begins where the previous one ended. This contrast is illustrated in Fig.~\ref{fig:window_quantized} in Appendix~\ref{appendix_windows}.} 
{We use $\tau=3\%$ for the analysis, as this choice also yields the lowest held-out error for power estimation in Section~\ref{subsec:power_validation}.} 
As the dataset is soil-dominated, the following analysis focuses on soil, with the detailed decomposed granite and sand analysis deferred to Appendix~\ref{appendix_granite_sand}.

\subsubsection{Collinearity Analysis and Regressor Clubbing}
\label{subsubsec:collin}

Table~\ref{tab:corr_soil} reports, for soil, the Pearson correlation matrix among the columns of the design matrix $D$. Its $(a,a')$ entry is the correlation between the duration vectors of subactivities $a$ and $a'$ across the $d$ windows formulated as
\small{
\begin{equation}
r_{a,a'} = \frac{\sum_{i=1}^{d}(D_{i,a}-\bar{D}_a)(D_{i,a'}-\bar{D}_{a'})}{\sqrt{\sum_{i=1}^{d}(D_{i,a}-\bar{D}_a)^2}\;\sqrt{\sum_{i=1}^{d}(D_{i,a'}-\bar{D}_{a'})^2}},
\label{eq:pearson_corr}
\end{equation}}%
\normalsize
where $\bar{D}_a=\tfrac{1}{d}\sum_{i} D_{i,a}$ is the mean duration of subactivity $a$.
Loading and Swinging show a strong
correlation, with correlation coefficient $r=0.81$. This is mechanically expected: loading and swinging form a single load-and-rotate cycle (fill, rotate to dump, unload, rotate back), so they are temporally interleaved and their per-window durations co-vary almost exactly. The
corresponding columns of $D$ are therefore nearly linearly dependent, and
the individual powers of Loading and Swinging are \emph{not separately
identifiable}. We resolve this by clubbing together the two collinear subactivities into the ``Loading$+$Swinging'' regressor. Among the estimated subactivities, every other positive power subactivity pair (idling is pinned to zero in the model in Section~\ref{subsubsec:estimation}) is at most weakly correlated
($|r|\le0.39$), so each is
individually identifiable and is retained as a separate regressor.

\begin{table}[t]
\centering
\caption{Pearson correlation score of subactivities. Loading and Swinging are strongly
correlated ($r=0.81$, bold).}
\vspace{-1 em}
\label{tab:corr_soil}
\footnotesize
\setlength{\tabcolsep}{4pt}
\begin{tabular}{@{}lrrrrr@{}}
\toprule
           & Digging    & Loading   & Swinging  & Traveling   & Idling   \\
\midrule
Digging    & $1.00$ & $-0.15$ & $-0.25$ & $-0.39$ & $-0.29$ \\
Loading    & $-0.15$ & $1.00$ & $\mathbf{0.81}$ & $-0.36$ & $-0.44$ \\
Swinging   & $-0.25$ & $\mathbf{0.81}$ & $1.00$ & $-0.17$ & $-0.28$ \\
Traveling & $-0.39$ & $-0.36$ & $-0.17$ & $1.00$ & $0.44$ \\
Idling     & $-0.29$ & $-0.44$ & $-0.28$ & $0.44$ & $1.00$ \\
\bottomrule
\end{tabular}
\end{table}

This
reduces the 5 raw subactivity classes for soil to $|\mathcal{A}|=4$ identifiable
regressors
\[
p = \big[\;p_{\mathrm{digging}}\; 
\underbrace{p_{\mathrm{ls}}}_{\text{Loading}+\text{Swinging}}\;
p_{\mathrm{traveling}}\;\;\; p_{\mathrm{idling}}\;\big]^{\!\top}.
\]

\subsubsection{Constrained Non-Negative Least Squares Estimation}\label{subsubsec:estimation}

Average subactivity power is physically non-negative, and a powered-on but
idle machine consumes negligible net energy relative to the working subactivities; we encode both facts as hard
constraints. The power estimation thus solves the following constrained, non-negative
least-squares program.

\begin{equation}
\begin{aligned}
\hat{p} \;=\; \arg\min_{p}\;\;
& \big\| D p - b \big\|_2^2 \\
\text{s.t.}\;\;
& p \succeq 0, \qquad p_{\mathrm{idling}} = 0.
\end{aligned}
\label{eq:nnls}
\end{equation}

\subsection{Subactivity Power Estimation Results and Validation}\label{subsec:power_validation}
Because the dataset is modest in size, we quantify both predictive
accuracy and coefficient stability by resampling rather than relying on
a single train/test split. The error analysis is performed via repeated $K$-fold cross-validation~\cite{wong2019reliable} with
$K=5$ and $R=200$ repetitions,\footnote{Repeated $K$-fold gives
uniform test coverage, i.e., every equation is held out exactly $R$
times, in contrast to the binomial coverage of random hold-out, so no equation is over or underrepresented in the test
statistics.} for a total of $KR=1000$ model
fits. In each fold, the estimator \eqref{eq:nnls} is fit on $80\%$ of the
data and evaluated on the held-out $20\%$. 
The full-data subactivity durations, and their cross-validated mean powers, standard deviation (std), and $95\%$ confidence intervals (CI) across the 1000 fits
are given in Table~\ref{tab:coeffs_soil}. The held-out error
metrics, i.e., comparing each fold's predictions $\hat{b}=D_{\mathrm{test}}\hat{p}$ against
the observed energies $b_{\mathrm{test}}$ are summarized in Table~\ref{tab:metrics_soil}. 

Three observations follow. First, the model predicts held-out
per-window energy to within roughly $0.13$~kWh, i.e., a normalized mean absolute error (NMAE) and mean absolute percentage error (MAPE) of
about $17\%$ and $18\%$, respectively. {A substantial part of this error reflects the resolution of the telematics and data limitation rather than model misfit. This is substantiated by the fact that a window's measured energy drop carries a worst case error of $\pm1\%$~SOC, which, for the smallest admissible windows ($\Delta\mathrm{SOC}=\tau=3\%$), is a worst-case relative error of $\approx33\%$.} 
Second, the error is stable
across resampling: the standard deviation across the $1000$ fits is
modest (e.g., MAE $0.130\pm0.041$~kWh), so the reported accuracy is not an
artifact of any particular split. Third, the subactivity powers are estimated with tight
and physically sensible confidence intervals: digging ($\approx4.8$~kW)
and traveling ($\approx4.7$~kW) are the highest, while loading$+$swinging
is the lowest at $\approx3.2$~kW. 

\begin{table}[t]
\centering
\scriptsize
\setlength{\tabcolsep}{2pt}
\renewcommand{\arraystretch}{1}
\caption{Estimated subactivity power for soil with $\tau=3\%$.}
\vspace{-0.8em}
\label{tab:coeffs_soil}
\begin{tabularx}{\columnwidth}{@{}lYYYY@{}}
\toprule
& Digging
& \makecell{Loading\\$+$Swinging}
& Traveling
& Idling \\
\midrule
\makecell[l]{Data share [\%]}
& 18
& 36
& 15
& 31 \\

\makecell[l]{Mean$\pm$std [kW]}
& $4.77\pm0.23$
& $3.18\pm0.23$
& $4.72\pm0.54$
& $0.00\pm0.00$ \\

\makecell[l]{95\% CI [kW]}
& $[4.30,\,5.20]$
& $[2.73,\,3.69]$
& $[3.59,\,5.86]$
& $[0.00,\,0.00]$ \\
\bottomrule
\end{tabularx}
\end{table}
\begin{table}[t]
\centering
\scriptsize
\setlength{\tabcolsep}{2pt}
\renewcommand{\arraystretch}{1}
\caption{Held-out energy prediction error.}
\vspace{-0.8em}
\label{tab:metrics_soil}
\begin{tabularx}{\columnwidth}{@{}YYYY@{}}
\toprule
MAE [kWh]
& RMSE [kWh]
& MAPE [\%]
& \makecell{Normalized\\MAE [\%]} \\
\midrule
$0.130\pm0.041$
& $0.156\pm0.050$
& $18.1\pm5.4$
& $17.4\pm5.7$ \\
\bottomrule
\end{tabularx}
\end{table}

\section{Optimal Scheduling Problem Formulation}
\label{sec:problem}
This section formulates the joint optimization of CEV work scheduling and MCS charging. Optimizing CEV work schedules is motivated by the power analysis in Section~\ref{sec:power_consumption}, which shows that subactivity power demands vary substantially. Thus, even with the same total work duration, reordering subactivities can reshape the CEV demand profile and hence when and how much MCSs charge from the grid.

We denote the scheduling-horizon granularity in hours by $\Delta T$ and the set of operating time intervals by $\mathcal{T}=\{1,2,\ldots,t_{\mathrm{end}}\}$, which indexes decision variables applied over each interval, such as charging and discharging powers from 01:00 to 01:15~h when $\Delta T=0.25$~h. $\mathcal{T}^{+}=\{0,1,\cdots,t_{\mathrm{end}}\}$ denotes the set of boundary time instants within the scheduling horizon. $t=0$ denotes the initial time instant of the scheduling horizon. To meet the charging demands of CEVs (denoted by set $\mathcal{E}$), we consider a network consisting of MCSs (denoted by set $\mathcal{M}$), depots for MCS charging called grid connection nodes (denoted by set $\mathcal{N}^{g}$), and construction nodes (denoted by set $\mathcal{N}^{c}$). {$\mathcal{N}=\mathcal{N}^{g} \cup \mathcal{N}^{c}$ denotes the set of all the nodes in the network.} The CEVs are assigned to fixed construction nodes, and are charged through the MCSs. Electric pickup trucks tow the MCSs between the nodes, such as between construction nodes or from a construction node to a grid connection node for recharging. 

\subsection{Objective Function}\label{subsec:objective}


The objective is to minimize the total operating cost over the scheduling horizon, and is formulated as

\small{
\begin{align}
\mathop{\text{minimize}}\limits_{\mathcal{D}}\, J
=& \sum_{m\in\mathcal{M}}\sum_{t\in\mathcal{T}}
\lambda_t^{\mathrm{elec}} P_{m,t}^{\mathrm{ch,tot}}\Delta T \nonumber \\
&+ \sum_{m\in\mathcal{M}}\sum_{t\in\mathcal{T}}
\lambda_t^{\mathrm{CO_2}} \lambda^{\mathrm{em}}
P_{m,t}^{\mathrm{ch,tot}}\Delta T  \nonumber \\
&+ \lambda^{\mathrm{NC}}P^{\mathrm{NC}}
+ \lambda^{\mathrm{OP}}P^{\mathrm{OP}}  \nonumber \\
&+\rho^{\mathrm{miss}}\sum_{i\in\mathcal{N}^{c}}\sum_{a\in\mathcal{A}}
s_{i,a}^{\mathrm{miss}} \nonumber \\
&+ \rho^{\mathrm{travel}}\Delta T\sum_{m \in \mathcal{M}}\sum_{i,j\in\mathcal{N}}\sum_{t\in\mathcal{T}}
y_{m,i,j,t}.\label{eq:objective}
\end{align}}
\normalsize

\noindent The first term in~\eqref{eq:objective} represents the energy cost of charging the MCSs from
the grid; with $\lambda_t^{\mathrm{elec}}$ and $P_{m,t}^{\mathrm{ch,tot}}$ denoting the electricity charge rate (\$/kWh) and total charging power (kW) of MCS $m$ from the grid over time interval $t$; respectively. The second term in~\eqref{eq:objective} represents the monetized carbon-emissions cost
associated with grid energy consumption; with $\lambda_t^{\mathrm{CO_2}}$ and $\lambda^{\mathrm{em}}$ denoting the carbon emissions intensity ($\mathrm{kgCO}_{2}$/kWh) and conversion factor of carbon emissions to dollar terms (\$/$\mathrm{kgCO}_{2}$); respectively. The third and fourth term in~\eqref{eq:objective} capture the non-coincident (NC) and on-peak (OP) demand charges. NC demand charges (NCDC) are the cost levied for the maximum power withdrawn from the grid in a month, while OP demand charges (OPDC) are the cost levied for the maximum power withdrawn from the grid in all the OP hours (4-9 pm) of the month. $\lambda^{\mathrm{NC}}$ and $\lambda^{\mathrm{OP}}$ denote the NCDC rate and OPDC rate (\$/kW); respectively, while $P^{\mathrm{NC}}=\max \{\sum_{m\in\mathcal{M}}P_{m,t}^{\mathrm{ch,tot}}\}_{t \in \mathcal{T}}$ and $P^{\mathrm{OP}}=\max \{\sum_{m\in\mathcal{M}}P_{m,t}^{\mathrm{ch,tot}}\}_{t \in \mathcal{T}^{\mathrm{op}}}$ denote the NC demand peak and OP demand peak; respectively.\footnote{Although for electricity bills, the NC demand peak and OP demand peak are computed over a month, in operation it is rare to schedule CEV/MCS over a month. Thus, in practice we penalize the peaks over the scheduling horizon with the intuition that repeatedly minimizing the peaks over rolling scheduling horizons acts as a surrogate to minimize the monthly peaks~\cite{ghosh2025adaptive,ghosh2025baseline}.} 

The fifth term in~\eqref{eq:objective} penalizes unmet construction work, where $s_{i,a}^{\mathrm{miss}}$ is the missed duration (hours) for subactivity $a$ at construction node $i$ over the entire scheduling horizon. $\rho^{\mathrm{miss}}$ is a conversion factor of missed work to dollar terms (\$/hour), based on construction crew labor, delay, and operational costs. Note that $\rho^{\mathrm{miss}}$ is chosen as a high number so that the penalty on the unmet work is an \emph{order of magnitude higher} than the other terms, as finishing construction work on time is the most crucial objective for construction operators. {This is because any leftover work that has to be completed beyond the allocated time entails additional equipment rental and labor costs as well as construction delays and associated penalties.} The last term in~\eqref{eq:objective} penalizes the labor cost for towing the MCSs from one place to another over the scheduling horizon; where $y_{m,i,j,t}$ is a binary variable which takes a value of 1 if the MCS $m$ is moving along path $(i,j)$ over time interval $t$, and is 0 otherwise, and $\rho^{\mathrm{travel}}$ is the labor cost (\$/hour) associated with towing the MCSs from one place to another. $\mathcal{D}$ represents the set of all the decision variables of the optimization (see nomenclature in Appendix~\ref{appendix_nomenclature}).

\subsection{Constraints}\label{subsec:constraints}

The constraints of the optimization problem are organized into four groups as presented below.

\subsubsection{Charging and Discharging Power Constraints of the MCSs and CEVs}\label{subsubsec:charge_discharge_constraints}

\small{
\begin{subequations}
\label{eq:power_aggregation}
\begin{align}
&P_{m,t}^{\mathrm{ch,tot}}
= \sum_{i\in\mathcal{N}^{g}} P_{m,i,t}^{\mathrm{ch,MCS}},
\quad \forall m\in\mathcal{M},\; t\in\mathcal{T},
\label{eq:mcs_total_charging}\\
&P_{m,t}^{\mathrm{dch,tot}}
= \sum_{i\in\mathcal{N}^{c}} P_{m,i,t}^{\mathrm{dch,MCS}},
\quad \forall m\in\mathcal{M},\; t\in\mathcal{T},
\label{eq:mcs_total_discharging}\\
&P_{m,i,t}^{\mathrm{dch,MCS}}
=0,
\quad \forall m\in\mathcal{M},\; i\in\mathcal{N}^{g},\; t\in\mathcal{T},
\label{eq:no_discharge_grid}\\
&P_{m,i,t}^{\mathrm{ch,MCS}}
=0,
\quad \forall m\in\mathcal{M},\; i\in\mathcal{N}^{c},\; t\in\mathcal{T},
\label{eq:no_charge_site}\\
&P_{m,i,t}^{\mathrm{dch,MCS}}
= \sum_{e\in\mathcal{E}}P_{m,i,e,t}^{\mathrm{MCS}\rightarrow\mathrm{CEV}},
\quad \forall m\in\mathcal{M},\; i\in\mathcal{N}^{c},\; t\in\mathcal{T}.
\label{eq:mcs_discharge_to_cev}\\
&P_{m,i,t}^{\mathrm{ch,MCS}}, P_{m,i,t}^{\mathrm{dch,MCS}},P_{m,i,e,t}^{\mathrm{MCS}\rightarrow\mathrm{CEV}} \geq 0, \nonumber\\
&\qquad \qquad \qquad \qquad \qquad \quad \forall m\in\mathcal{M},\; i\in\mathcal{N},\; e \in \mathcal{E},\; t\in\mathcal{T}.
\label{eq:pos_power}
\end{align}
\end{subequations}}
\normalsize

{Constraints \eqref{eq:mcs_total_charging} and \eqref{eq:mcs_total_discharging}
define the total charging and discharging powers of each MCS $m$ over time interval $t$ as a sum of node specific charging and discharging powers, respectively. $P_{m,t}^{\mathrm{ch,tot}}/P_{m,t}^{\mathrm{dch,tot}}$ denote the total charging/discharging power of MCS $m$ over time interval $t$, and $P_{m,i,t}^{\mathrm{ch,MCS}} / P_{m,i,t}^{\mathrm{dch,MCS}}$ denote the charging/discharging power of MCS $m$ at node $i$ over time interval $t$.} {Note that each MCS can be present at only one node during each time interval, and is described later in constraint \eqref{eq:one_location_or_travel}.} 
Constraints
\eqref{eq:no_discharge_grid} and \eqref{eq:no_charge_site} enforce that
MCSs charge only at grid nodes and discharge only at construction nodes; respectively. Constraint \eqref{eq:mcs_discharge_to_cev} ensures that the discharging power
of an MCS $m$ at construction node $i$ over time interval $t$ equals the total power transferred from
that MCS to all the connected CEVs, where $P_{m,i,e,t}^{\mathrm{MCS}\rightarrow\mathrm{CEV}}$ denotes the power transferred by MCS $m$, at node $i$, to CEV $e$ over time interval $t$. {The connection status of MCS $m$ at node $i$ to CEV $e$ over time interval $t$ is denoted by a binary variable $\rho_{m,i,e,t}$ and is described later in~\eqref{eq:plug_power_limit}.} Constraint~\eqref{eq:pos_power} ensures that all the charging and discharging powers of the MCS, and the power transfer from the MCS to CEVs are non-negative. 

\small{
\begin{subequations}
\label{eq:mcs_power_limits}
\begin{align}
&P_{m,i,t}^{\mathrm{ch,MCS}}
\le \text{CH}_m^{\mathrm{MCS}} z_{m,i,t},\;
 \forall m\in\mathcal{M},\; i\in\mathcal{N}^g,\; t\in\mathcal{T},
\label{eq:mcs_charging_limit}\\
&P_{m,i,t}^{\mathrm{dch,MCS}}
\le \text{DCH}_m^{\mathrm{MCS}} z_{m,i,t},\;
 \forall m\in\mathcal{M},\; i\in\mathcal{N}^c,\; t\in\mathcal{T}.
\label{eq:mcs_discharging_limit}
\end{align}
\end{subequations}}
\normalsize

\noindent Here $z_{m,i,t}$ is a binary variable, where $z_{m,i,t}=1$ if and only if (iff) MCS~$m$ is present at node~$i$ over interval~$t$ ($0$ otherwise). Constraint~\eqref{eq:mcs_charging_limit} permits grid charging only when $z_{m,i,t}=1$ at a grid charging node, and caps it at the MCS charging capacity $\text{CH}_m^{\mathrm{MCS}}$. 
Constraint~\eqref{eq:mcs_discharging_limit} analogously limits the discharging power at each construction node to the discharging capacity $\text{DCH}_m^{\mathrm{MCS}}$, only when the MCS is present ($z_{m,i,t}=1$). 

\small{
\begin{subequations}
\label{eq:cev_charging_limits}
\begin{align}
&P_{m,i,e,t}^{\mathrm{MCS}\rightarrow\mathrm{CEV}}
\le \text{DCH}_m^{\mathrm{plug}}\rho_{m,i,e,t}, \nonumber \\
&\qquad \qquad \qquad \qquad \forall m\in\mathcal{M},\; i\in\mathcal{N}^c,\; e\in\mathcal{E},\; t\in\mathcal{T},
\label{eq:plug_power_limit}\\
&\sum_{m\in\mathcal{M}}P_{m,i,e,t}^{\mathrm{MCS}\rightarrow\mathrm{CEV}}
\le \text{CH}_e^{\mathrm{CEV}}\mu_{i,e,t},
 \forall i\in\mathcal{N}^{c},\; e\in\mathcal{E},\; t\in\mathcal{T},
\label{eq:cev_acceptance_limit}\\
&\mu_{i,e,t}= \sum_{m \in \mathcal{M}}\rho_{m,i,e,t}, \quad
 \forall i\in\mathcal{N}^{c},\; e\in\mathcal{E},\; t\in\mathcal{T},
\label{eq:cev_acceptance_limit_2}\\
&\sum_{e\in\mathcal{E}}\rho_{m,i,e,t}
\le C_m^{\mathrm{plug}},
\quad \forall m\in\mathcal{M},\; i\in\mathcal{N}^{c},\; t\in\mathcal{T}.
\label{eq:plug_count_limit}
\end{align}
\end{subequations}}
\normalsize

\noindent Constraint \eqref{eq:plug_power_limit} limits the power delivered through
each MCS plug to $\text{DCH}_m^{\mathrm{plug}}$, and forces the MCS-to-CEV power to zero unless the binary connection variable is active, where $\text{DCH}_m^{\mathrm{plug}}$ denotes the discharging capacity of plugs of MCS $m$, and $\rho_{m,i,e,t}$ denotes the connection status of MCS $m$ at node $i$ to CEV $e$ over time interval $t$. 
Constraint \eqref{eq:cev_acceptance_limit}
caps the charging acceptance rate of each CEV below $\text{CH}_e^{\mathrm{CEV}}$ and permits charging iff the binary variable $\mu_{i,e,t}=1$, indicating the CEV $e$ is ready to accept charge at node $i$ over time interval $t$. 
{Constraint~\eqref{eq:cev_acceptance_limit_2} links the two indicators, setting the CEV-side acceptance variable $\mu_{i,e,t}$ equal to the sum of the MCS-side connection variables $\rho_{m,i,e,t}$ over all MCSs, so that $\mu_{i,e,t}=1$ precisely when some MCS is plugged into CEV $e$ at node $i$ during interval $t$.} Constraint \eqref{eq:plug_count_limit} limits the number
of CEVs simultaneously connected to a given MCS by the number of outlet plugs from the MCS, given by $C_m^{\mathrm{plug}}$.


\subsubsection{Energy Consumption and State-of-Energy (SOE) Constraints of the MCSs and CEVs}
\label{subsubsec:soe_constraints}

\small{
\begin{subequations}
\label{eq:work_power_constraints}
\begin{align}
&u_{i,e,t,a}
\le A_{i,e},\quad
 \forall i\in\mathcal{N}^{c},\; e\in\mathcal{E},\; t\in\mathcal{T}, a\in \mathcal{A},
\label{eq:combined_work_limit}\\
&u_{i,e,t,a} = 0, \quad \forall i \in \mathcal{N}^c, e\in\mathcal{E},\; t\in\mathcal{T}\setminus\{\mathcal{T}^{\mathrm{work}}\}, a \in \mathcal{A}, \label{eq:working_hour_limit}\\
&\sum_{a\in\mathcal{A}}u_{i,e,t,a}+\mu_{i,e,t}
\le 1,\quad
 \forall i\in\mathcal{N}^{c},\; e\in\mathcal{E},\; t\in\mathcal{T},
\label{eq:work_charge_exclusive_binary}\\
&P_{i,e,t}^{\mathrm{work}}
= \sum_{a\in\mathcal{A}}p_a u_{i,e,t,a},\quad
 \forall i\in\mathcal{N}^{c},\; e\in\mathcal{E},\; t\in\mathcal{T}.
\label{eq:work_power_activity}
\end{align}
\end{subequations}}
\normalsize

\noindent The above constraints couple the charging decisions of the CEVs with their work schedule and construction node assignment.
{Constraint~\eqref{eq:combined_work_limit}
limits the CEV to only work in pre-specified construction nodes it is assigned to, where $u_{i,e,t,a}$ is a binary variable taking the value 1 iff CEV $e$ is performing subactivity $a$ at node $i$ over time interval $t$, and the assignment matrix is denoted by $A_{i,e}$ which takes a value of 1 iff CEV $e$ is assigned to node $i$, and is 0 otherwise. $u_{i,e,t,a}$ takes a value of 0 for hours other than the pre-specified working hours as formulated in~\eqref{eq:working_hour_limit}.} Constraint
\eqref{eq:work_charge_exclusive_binary} enforces that a CEV cannot simultaneously charge
and work over the same time interval $t$. Constraint
\eqref{eq:work_power_activity} defines the work power $P_{i,e,t}^{\mathrm{work}}$ of CEV $e$ at node $i$ over time interval $t$ from the performed
construction subactivity, where $p_a$ is the mean power consumption of the CEV while performing subactivity $a$.

\small{
\begin{subequations}
\label{eq:soe_dynamics}
\begin{align}
&\mathrm{SOE}_{m,t+1}^{\mathrm{MCS}}
= \;\mathrm{SOE}_{m,t}^{\mathrm{MCS}}
+ \eta_m P_{m,t+1}^{\mathrm{ch,tot}}\Delta T
- \frac{P_{m,t+1}^{\mathrm{dch,tot}}\Delta T}{\eta_m}, \nonumber \\
&\qquad \qquad \qquad
\qquad \qquad \forall m\in\mathcal{M},\; t\in\mathcal{T}^{+}\setminus\{t_{\mathrm{end}}\},
\label{eq:mcs_soe_dynamics}\\
&\mathrm{SOE}_{e,t+1}^{\mathrm{CEV}}
=\; \mathrm{SOE}_{e,t}^{\mathrm{CEV}}
+ {\eta}_{e}\sum_{m\in\mathcal{M}}\sum_{i\in\mathcal{N}^{c}}
P_{m,i,e,t+1}^{\mathrm{MCS}\rightarrow\mathrm{CEV}}\Delta T \nonumber \\
&\qquad \qquad  \qquad- \sum_{i\in\mathcal{N}^{c}}P_{i,e,t+1}^{\mathrm{work}}\Delta T,
\quad  \forall e\in\mathcal{E},\; t\in\mathcal{T}^{+}\setminus\{t_{\mathrm{end}}\}.
\label{eq:cev_soe_dynamics}
\end{align}
\end{subequations}}
\normalsize

\noindent Equation~\eqref{eq:mcs_soe_dynamics} updates the MCS state-of-energy (SOE) by adding grid-charged
energy, and subtracting MCS-to-CEV discharged energy after efficiency losses, where $\eta_m$ is the charging/discharging efficiency of MCS $m$. 
Equation \eqref{eq:cev_soe_dynamics} updates the CEV SOE by adding energy received from MCSs after efficiency losses and subtracting energy consumed for work, where $\eta_e$ is the charging efficiency of CEV $e$. 

\small{\setlength{\abovedisplayskip}{1.5pt}
\setlength{\belowdisplayskip}{1.5pt}
\setlength{\abovedisplayshortskip}{0pt}
\setlength{\belowdisplayshortskip}{0pt}
\setlength{\jot}{2pt}
\begin{subequations}
\label{eq:soe_boundary_bounds}
\begin{align}
&\mathrm{SOE}_{m,0}^{\mathrm{MCS}}
= \mathrm{SOE}_{m}^{\mathrm{MCS,ini}} = \mathrm{SOE}_{m,t_{\mathrm{end}}}^{\mathrm{MCS}},
&& \forall m\in\mathcal{M},
\label{eq:mcs_initial_soe}\\
&\mathrm{SOE}_{e,0}^{\mathrm{CEV}}
= \mathrm{SOE}_{e}^{\mathrm{CEV,ini}}=\mathrm{SOE}_{e,t_{\mathrm{end}}}^{\mathrm{CEV}},
&& \forall e\in\mathcal{E},
\label{eq:cev_initial_soe}\\
&\mathrm{SOE}_{m}^{\mathrm{MCS,min}}
\le \mathrm{SOE}_{m,t}^{\mathrm{MCS}}
\le \mathrm{SOE}_{m}^{\mathrm{MCS,max}},
&& \forall m\in\mathcal{M},\; t\in\mathcal{T}^{+},
\label{eq:mcs_soe_bounds}\\
&\mathrm{SOE}_{e}^{\mathrm{CEV,min}}
\le \mathrm{SOE}_{e,t}^{\mathrm{CEV}}
\le \mathrm{SOE}_{e}^{\mathrm{CEV,max}},
&& \forall e\in\mathcal{E},\; t\in\mathcal{T}^{+}.
\label{eq:cev_soe_bounds}
\end{align}
\end{subequations}}
\normalsize

\noindent Equations \eqref{eq:mcs_initial_soe}--\eqref{eq:cev_initial_soe} set the
initial and final SOE of all MCSs and CEVs to be equal to impose an energy-neutral terminal condition, preventing the optimizer from artificially depleting batteries at the end of
the horizon, and be ready for next day's work. Constraints \eqref{eq:mcs_soe_bounds}--\eqref{eq:cev_soe_bounds}
enforce the allowable SOE operating ranges for the MCS and CEV; respectively.


\subsubsection{Spatial Routing and Connection Status of the MCSs and CEVs}\label{subsubsec:routing_constraints}

\small{
\begin{subequations}
\label{eq:connection_status}
\begin{align}
\rho_{m,i,e,t}
&\le A_{i,e},
&& \forall m\in\mathcal{M},\; i\in\mathcal{N}^c,\; e\in\mathcal{E},\; t\in\mathcal{T},
\label{eq:rho_assignment}\\
\rho_{m,i,e,t}
&\le z_{m,i,t},
&& \forall m\in\mathcal{M},\; i\in\mathcal{N}^c,\; e\in\mathcal{E},\; t\in\mathcal{T}.
\label{eq:rho_location}
\end{align}
\end{subequations}}
\normalsize

\noindent Constraint \eqref{eq:rho_assignment} allows MCS $m$ to connect to CEV $e$ at node $i$ 
iff the CEV $e$ is assigned to that node, through the assignment matrix. Similarly, constraint \eqref{eq:rho_location} allows the MCS $m$ to connect to a CEV at node $i$ iff the MCS $m$ is physically present at that node. {Recall that $\rho_{m,i,e,t}$ is the binary MCS--CEV connection variable ($1$ iff MCS $m$ at node $i$ is plugged into CEV $e$ during interval $t$), $A_{i,e}$ the assignment matrix ($1$ iff CEV $e$ is assigned to node $i$), and $z_{m,i,t}$ the binary MCS presence variable ($1$ iff MCS $m$ is present at node $i$ during interval $t$).} 

\small{
\begin{subequations}
\label{eq:routing_basic}
\begin{align}
&x_{m,i,i,t}
=0,
\quad \forall m\in\mathcal{M},\; i\in\mathcal{N},\; t\in\mathcal{T},
\label{eq:no_self_loop}\\
& y_{m,i,j,t} = \sum_{\tau=\max\{1,\;t-\tau_{i,j}^{\mathrm{trv}}+1\}}^{t} x_{m,i,j,\tau}, \nonumber \\
&\qquad \qquad \qquad \qquad \forall m\in\mathcal{M},\; i,j\in\mathcal{N},\; i\ne j,\; t\in\mathcal{T} \label{eq:y_trv_definition} \\
&\sum_{i\in\mathcal{N}}z_{m,i,t}
+\sum_{\substack{i,j\in\mathcal{N}\\i\ne j}}y_{m,i,j,t}
= 1,
\quad \forall m\in\mathcal{M},\; t\in\mathcal{T},
\label{eq:one_location_or_travel}
\end{align}
\end{subequations}}
\normalsize

\noindent Constraint \eqref{eq:no_self_loop} eliminates travel from a node to itself, where $x_{m,i,j,t}$ is a binary variable that takes a value of 1 iff MCS $m$ departs node $i$ towards node $j$ (along path $(i,j)$) over time interval $t$.  {Let $\tau_{i,j}^{\mathrm{trv}}$ denote the number of time intervals (each of length $\Delta T$) an MCS takes to travel from node $i$ to node $j$, which is a user input fixed by the network topology.} 
{Constraint~\eqref{eq:y_trv_definition} sets the binary travel indicator $y_{m,i,j,t}$ to $1$ iff MCS $m$ is in transit from node $i$ to node $j$ over time interval $t$. Since a trip on path $(i,j)$ lasts $\tau_{i,j}^{\mathrm{trv}}$ intervals, the MCS is in transit at $t$ iff it departed node $i$ for node $j$ over some time interval $\tau$ with $t-\tau_{i,j}^{\mathrm{trv}}+1 \le \tau \le t$, which is the range the summation in~\eqref{eq:y_trv_definition} covers. The $\max\{1,\cdot\}$ simply prevents this look-back window from extending before the first time interval of the horizon.} 
Constraint \eqref{eq:one_location_or_travel} ensures that an MCS is either parked at one node or traveling on one route during a time interval.

\small{
\begin{subequations}
\label{eq:arrival_departure}
\begin{align}
&\beta_{m,i,t}^{\mathrm{dep}}
= \sum_{\substack{j\in\mathcal{N}\\j\ne i}}x_{m,i,j,t},
\quad \forall m\in\mathcal{M},\; i\in\mathcal{N},\; t\in\mathcal{T},
\label{eq:departure_def}\\
&\beta_{m,j,t}^{\mathrm{arr}}
= \sum_{\substack{i\in\mathcal{N}\\i\ne j\\t-\tau_{i,j}^{\mathrm{trv}}\in\mathcal{T}}}
x_{m,i,j,t-\tau_{i,j}^{\mathrm{trv}}},
\quad \forall m\in\mathcal{M},\; j\in\mathcal{N},\; t\in\mathcal{T},
\label{eq:arrival_def}\\
&\beta_{m,i,t}^{\mathrm{arr}}-\beta_{m,i,t}^{\mathrm{dep}}
= z_{m,i,t}-z_{m,i,t-1}, \nonumber \\
&\qquad \qquad \qquad \qquad  \forall m\in\mathcal{M},\; i\in\mathcal{N},\;
t\in\mathcal{T}\setminus\{1\},
\label{eq:location_transition}\\
&\sum_{t\in\mathcal{T}}\beta_{m,i,t}^{\mathrm{arr}}
= \sum_{t\in\mathcal{T}}\beta_{m,i,t}^{\mathrm{dep}},
\quad \forall m\in\mathcal{M},\; i\in\mathcal{N},
\label{eq:flow_conservation}\\
&\beta_{m,i,t}^{\mathrm{arr}}+\beta_{m,i,t}^{\mathrm{dep}}
\le 1,
\quad \forall m\in\mathcal{M},\; i\in\mathcal{N},\; t\in\mathcal{T}.
\label{eq:no_arrive_depart_same_time}
\end{align}
\end{subequations}}
\normalsize

\noindent Constraint \eqref{eq:departure_def} defines the binary departure indicator of an MCS, $\beta_{m,i,t}^{\mathrm{dep}}$, which takes a value of 1 iff MCS $m$ departs node $i$ along some outgoing node $j\neq i$ over time interval $t$. {Constraint \eqref{eq:arrival_def} defines the binary arrival indicator of an MCS, $\beta_{m,j,t}^{\mathrm{arr}}$, which takes a value of 1 iff MCS $m$ arrives at node $j$ over time interval $t$: an arrival at node $j$ over time interval $t$ corresponds to a departure from some node $i$ toward $j$ exactly $\tau_{i,j}^{\mathrm{trv}}$ intervals earlier, hence the time-shifted term $x_{m,i,j,\,t-\tau_{i,j}^{\mathrm{trv}}}$.} Constraint \eqref{eq:location_transition} updates the MCS node-presence
variable $z_{m,i,t}$ using arrivals and departures. Constraint
\eqref{eq:flow_conservation} balances total arrivals and departures over the
horizon, while \eqref{eq:no_arrive_depart_same_time} prevents an MCS from
arriving at and departing from the same node in the same interval.

\subsubsection{Operational Adjustment for CEV Work Scheduling}\label{subsubsec:operational_constraints}

Here, we present a distinction between all the CEV subactivities in $\mathcal{A}$, and productive construction related subactivities in $\mathcal{A}^{\mathrm{con}}$ (only digging, and loading+swinging) with $\mathcal{A}^{\mathrm{con}} \subseteq \mathcal{A}$, to model the CEV working constraints.

\small{
\begin{subequations}
\label{eq:work_scheduling}
\begin{align}
&\sum_{a\in\mathcal{A}}u_{i,e,t,a}
\le 1,
\quad \forall i\in\mathcal{N}^{c},e\in\mathcal{E},\; t\in\mathcal{T},
\label{eq:one_activity}\\
&\Delta T\sum_{e\in\mathcal{E}}\sum_{t\in\mathcal{T}}u_{i,e,t,a}
+s_{i,a}^{\mathrm{miss}}
= H_{i,a},
\; \forall i\in\mathcal{N}^{c},\; a\in\mathcal{A}^{\mathrm{con}},
\label{eq:activity_requirement}\\
&\gamma_{a\rightarrow a'}\sum_{e \in \mathcal{E}}\sum_{\tau=1}^{t}u_{i,e,\tau,a'}
\le \kappa^{\mathrm{seq}}_{a \rightarrow a'}\sum_{e \in \mathcal{E}}\sum_{\tau=1}^{t}u_{i,e,\tau,a}, \nonumber \\
& \qquad \qquad \qquad \quad \forall i\in\mathcal{N}^{c},\; a,a' \in \mathcal{A},\;\; t\in\mathcal{T},
\label{eq:precedence}\\
&\sum_{\tau=t}^{t+t_{\mathrm{limit}}} \sum_{a\in\mathcal{A}} u_{i,e,\tau,a} \le t_{\mathrm{limit}},
\nonumber \\
&\qquad \forall i\in\mathcal{N}^{c},\; e\in\mathcal{E},\; t\in \{1,2,3,\ldots,t_{\mathrm{end}}-t_{\mathrm{limit}}\}.
\label{eq:cev_rest_rule}\\
&\kappa^{\mathrm{wt}}\sum_{\tau=1}^{t} u_{i,e,\tau,a^{\mathrm{trv}}}
\le
\sum_{a\in\mathcal{A}^{\mathrm{con}}}\sum_{\tau=1}^{t} u_{i,e,\tau,a}, \nonumber\\
&\qquad\qquad\qquad\qquad\qquad\quad \forall i\in\mathcal{N}^{c},  e\in\mathcal{E},\; t\in\mathcal{T},
\label{eq:travel_ub}\\
&\kappa^{\mathrm{wt}}\sum_{\tau=1}^{t} u_{i,e,\tau,a^{\mathrm{trv}}}
\ge
\sum_{a\in\mathcal{A}^{\mathrm{con}}}\sum_{\tau=1}^{t} u_{i,e,\tau,a} - \kappa^{\mathrm{wt}}, \nonumber\\
&\qquad\qquad\qquad\qquad\qquad\quad \forall i\in\mathcal{N}^{c}, e\in\mathcal{E},\; t\in\mathcal{T}.
\label{eq:travel_lb}
\end{align}
\end{subequations}}
\normalsize

\noindent Constraint \eqref{eq:one_activity} allows each CEV to perform at most one
subactivity over each time interval at each site. 
Constraint~\eqref{eq:activity_requirement} enforces the required productive work duration
$H_{i,a}$ at each site which is a user input, with slack $s_{i,a}^{\mathrm{miss}}$ capturing
unmet work in hours. {Constraint \eqref{eq:precedence} imposes precedence between subactivity pairs at each site through two user inputs: $\gamma_{a\rightarrow a'}\in\{0,1\}$, which equals $1$ iff subactivity $a$ must precede subactivity $a'$, and $\kappa^{\mathrm{seq}}_{a\rightarrow a'}\in\mathbb{N}$, which sets how many intervals (each of length $\Delta T$) of $a'$ each completed interval of $a$ unlocks. For example, with $a=$ digging and $a'=$ loading$+$swinging, loading$+$swinging can never run ahead of the digging completed so far, as material cannot be loaded before it has been dug, ensuring the physical consistency of the construction workflow. When $\gamma_{a\rightarrow a'}=0$, the left-hand side vanishes and the constraint is inactive.}
Constraint \eqref{eq:cev_rest_rule} imposes an operational rest requirement
on each CEV. {Specifically, over any $t_{\mathrm{limit}}+1$ consecutive intervals (each of length $\Delta T$), a CEV
is allowed to perform subactivities (digging, loading+swinging, or traveling) 
in at most $t_{\mathrm{limit}}$ intervals, where $t_{\mathrm{limit}}\in\mathbb{N}$ is a user input. Therefore,
each CEV must remain idle for at least one interval in every rolling $t_{\mathrm{limit}}+1$ interval window}, preventing uninterrupted operation and capturing practical operator-break requirements.

\noindent Constraints~\eqref{eq:travel_ub} and~\eqref{eq:travel_lb} couple the CEV's repositioning to its productive work, where $a^{\mathrm{trv}} $ denotes the traveling subactivity, and $\kappa^{\mathrm{wt}}\in \mathbb{N}$ is a user input which denotes the number of intervals of productive work between each travel interval. The upper bound~\eqref{eq:travel_ub} prevents more than one travel for every $\kappa^{\mathrm{wt}}$ productive work intervals (avoiding spurious repositioning), while the lower bound~\eqref{eq:travel_lb} forces at least one travel within every $\kappa^{\mathrm{wt}}$ productive work intervals (so work cannot proceed without repositioning). Together they yield an evenly interspersed productive work--travel pattern, consistent with the observed construction demonstration.

\section{Case study}\label{sec:case_study}
\begin{figure}[t]
    \centering
    \includegraphics[width=0.9\linewidth]{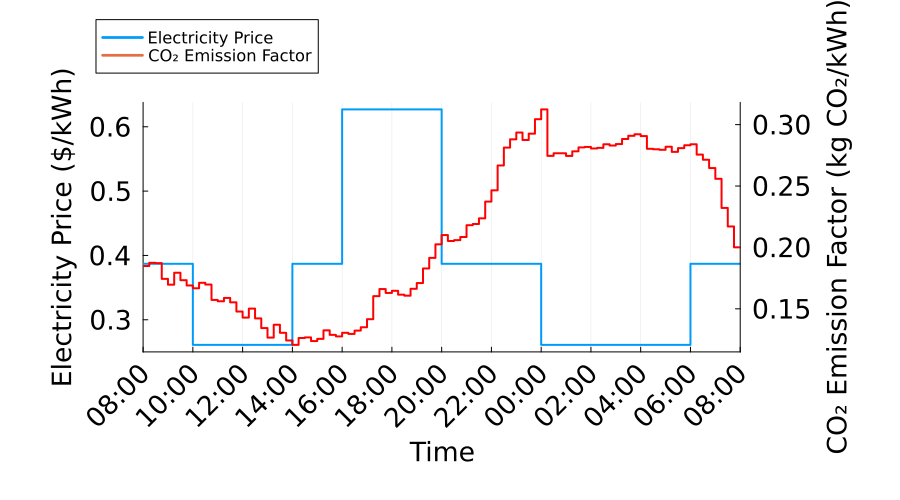}
    \vspace{-1.5em}
    \caption{Time-varying electricity charge rate $\lambda_t^{\mathrm{elec}}$ and grid carbon-emissions intensity $\lambda_t^{\mathrm{CO_2}}$ over the scheduling horizon.}
    \label{fig:electricity_carbon}
\end{figure}

In this section, we first introduce the case study scenarios and then lay out the baseline strategies against which we benchmark our proposed optimal MCS-CEV scheduling strategy. The design specifications of the CEV and MCS along with the operational parameters of the case study are given in Table~\ref{tab:specifications} in Appendix~\ref{appendix_case_study_settings}, and are based on the real-life construction setup. The electricity charge rate and carbon emissions intensity (from CAISO~\cite{caiso}) are time varying, and shown in Fig.~\ref{fig:electricity_carbon}. The demand charge rates are $\lambda^{\mathrm{NC}}=\$ 20.12$/kW, and $\lambda^{\mathrm{OP}}=\$ 20.58$/kW, according to SDG\&E's A6-TOU Transmission rates~\cite{sdge}.

\subsection{Scenarios} 
\label{subsec:scenario}
We consider four scenarios of increasing scale and operational complexity, varying the network topology and the numbers of CEVs and MCSs, as shown in Fig.~\ref{fig:topology} in Appendix~\ref{appendix_case_study_settings}: 1) \emph{Scenario~1} has one construction node with one CEV and one grid connection node hosting one MCS. Although the simplest scenario, it closely mirrors the UC San Diego field demonstration and is best suited for interpreting the optimization results. 2) \emph{Scenario 2} extends Scenario 1 to two CEVs at the same construction node, while retaining one MCS and one grid connection node. 3) \emph{Scenario~3} has two construction nodes, each with one CEV, served by one MCS charging at a single grid connection node. It highlights limited MCS availability, since both construction sites cannot be served simultaneously. 4) \emph{Scenario~4}, the largest case, has two construction nodes, each with two CEVs, and two MCSs charging at a single grid connection node.

In all scenarios, working hours are 8 am -- 12 noon, followed by a lunch break, and then again from 2 pm -- 5 pm. At each construction node, the productive work required is 3 hours of digging and 1.5 hours of loading+swinging. These settings are based on the field demonstration but can be adjusted to other operating schedules and workloads.

\begin{table*}[t]
\centering
\caption{Solver statistics, cost breakdown, missed work, MCS grid energy, and CO$_2$ emissions for the proposed strategy (Prop.) and baselines B1 and B2 across four scenarios. Bold values indicate the best solver statistics and primary performance outcomes within each scenario. The symbol $^{*}$ denotes runs that reached the one-hour time limit before proving optimality.}
\vspace{-1em}
\label{tab:cost_breakdown}
\footnotesize
\setlength{\tabcolsep}{6pt}
\renewcommand{\arraystretch}{0.8}
\setlength{\aboverulesep}{0.3ex}
\setlength{\belowrulesep}{0.4ex}

\begin{tabular}{@{}c l ccc !{\color{black!30}\vrule} ccc
!{\color{black!30}\vrule} ccc !{\color{black!30}\vrule} ccc@{}}
\toprule
 & &
\multicolumn{3}{c}{Scenario 1} &
\multicolumn{3}{!{\color{black!30}\vrule}c}{Scenario 2} &
\multicolumn{3}{!{\color{black!30}\vrule}c}{Scenario 3} &
\multicolumn{3}{!{\color{black!30}\vrule}c}{Scenario 4} \\
\cmidrule(lr){3-5}
\cmidrule(lr){6-8}
\cmidrule(lr){9-11}
\cmidrule(lr){12-14}

 & &
{Prop.} & B1 & B2 &
{Prop.} & B1 & B2 &
{Prop.} & B1 & B2 &
{Prop.} & B1 & B2 \\
\midrule

 & \textbf{MIP gap} [\%]
 & $\mathbf{0.0}$ & $\mathbf{0.0}$ & $\mathbf{0.0}$
 & $\mathbf{0.0}$ & $\mathbf{0.0}$ & $\mathbf{0.0}$
 & $\mathbf{0.0}$ & $\mathbf{0.0}$ & $\mathbf{0.0}$
 & $11.9$ & $\mathbf{0.0}$ & $\mathbf{0.0}$ \\

 & \textbf{Solve time} [s]
 & $28$ & $\mathbf{0.3}$ & $6$
 & $23$ & $\mathbf{0.3}$ & $4$
 & $327$ & $\mathbf{0.3}$ & $33$
 & $3600^{*}$ & $\mathbf{1.6}$ & $470$ \\

 & \textbf{Total cost [\$]}
 & $\mathbf{68}$ & $568$ & $566$
 & $\mathbf{57}$ & $66$ & $61$
 & $\mathbf{160}$ & $4116$ & $4102$
 & $\mathbf{110}$ & $1108$ & $1101$ \\

\midrule
\multirow{6}{*}{\rotatebox[origin=c]{90}{\makecell{Cost\\components}}}
 & NCDC [\$]
 & $39$ & $49$ & $37$
 & $38$ & $47$ & $33$
 & $102$ & $73$ & $59$
 & $62$ & $71$ & $65$ \\

 & OPDC [\$]
 & $0$ & $0$ & $0$
 & $0$ & $0$ & $0$
 & $0$ & $0$ & $0$
 & $0$ & $0$ & $0$ \\

 & Energy costs [\$]
 & $9.0$ & $8.5$ & $8.6$
 & $8.6$ & $8.4$ & $8.3$
 & $17.7$ & $12.9$ & $13.1$
 & $16.5$ & $15.8$ & $15.5$ \\

 & Carbon costs [\$]
 & $0.33$ & $0.36$ & $0.32$
 & $0.32$ & $0.35$ & $0.30$
 & $0.75$ & $0.55$ & $0.50$
 & $0.60$ & $0.59$ & $0.57$ \\

 & Missed-work penalty [\$]
 & ${0}$ & $500$ & $500$
 & ${0}$ & ${0}$ & ${0}$
 & ${0}$ & $4000$ & $4000$
 & ${0}$ & $1000$ & $1000$ \\

 & Travel costs [\$]
 & $20$ & $10$ & $20$
 & $10$ & $10$ & $20$
 & $40$ & $30$ & $30$
 & $30$ & $20$ & $20$ \\

\midrule
 & {Missed work [h]}
 & ${0}$ & $0.25$ & $0.25$
 & ${0}$ & ${0}$ & ${0}$
 & ${0}$ & $2.0$ & $2.0$
 & ${0}$ & $0.5$ & $0.5$ \\

 & MCS grid-charging energy [kWh]
 & $27.8$ & $26.8$ & $26.8$
 & $26.4$ & $26.4$ & $26.4$
 & $55.5$ & $41.7$ & $41.7$
 & $52.8$ & $48.6$ & $48.6$ \\

 & CO$_2$ emissions [kg]
 & $6.6$ & $7.3$ & $6.3$
 & $6.4$ & $7.1$ & $6.0$
 & $15.1$ & $10.9$ & $10.0$
 & $11.9$ & $11.8$ & $11.4$ \\

 & CO$_2$ per productive work-hour [kg/h]
 & ${1.47}$ & $1.72$ & $1.49$
 & $1.41$ & $1.57$ & ${1.34}$
 & $1.68$ & $1.56$ & ${1.43}$
 & ${1.32}$ & $1.39$ & $1.34$ \\
\bottomrule
\vspace{-2em}
\end{tabular}
\end{table*}

\subsection{Comparative Baseline Strategies}
\label{subsec:baseline_strategy}

To quantify the benefit of jointly optimizing the MCS and CEV schedules, we compare the proposed framework with two baselines, Baseline~1 (B1) and Baseline~2 (B2). In both baselines, part or all of the CEV schedule is fixed \emph{a priori}, while the remaining CEV and MCS decisions are optimized. This reflects the conventional MCS-scheduling paradigm, in which the CEV demand profile is treated as an exogenous input. In our setting, however, the CEV work schedule is jointly optimized with the charging decisions, subject to the routing and operational constraints in Sections~\ref{subsubsec:routing_constraints}--\ref{subsubsec:operational_constraints}. The comparison therefore isolates the value of co-optimizing the CEV schedule with the MCS schedule.

\emph{Reference CEV schedule generation:}
The fixed CEV profiles used by the baselines are generated by a rule-based reference scheduler that represents conventional construction operation. Its ``greedy'' objective (i)~minimizes unmet construction work through the missed-work penalty $\rho^{\mathrm{miss}}$, as finishing work on time is the most crucial construction parameter (also present in~\eqref{eq:objective}); (ii)~drives both CEV operation and MCS grid charging to the earliest feasible time intervals via time-weighted penalties; and (iii)~penalizes MCS in-transit time through $\rho^{\mathrm{travel}}$ as that reflects additional labor (also present in~\eqref{eq:objective}). In addition to the constraints in Section~\ref{subsec:constraints}, the reference scheduler imposes a ``work-to-minimum/charge-to-full'' duty cycle on every CEV wherever possible: the vehicle operates until its SOE is drawn down as close as possible to its lower limit and only then begins to charge, and once charging starts it must proceed uninterrupted until the battery is full. 
The resulting work-activity indicators $u_{i,e,t,a}$ and charging-mode indicators $\mu_{i,e,t}$ define the reference CEV schedule.

The baselines are described as follows:

1) \emph{Baseline~1} fixes both the CEV work schedule $u_{i,e,t,a}$ and charging intervals $\mu_{i,e,t}$ to the reference schedule. The optimization therefore determines only the MCS routing, MCS grid-charging schedule, and charging power delivered during the prescribed CEV charging intervals by minimizing objective~\eqref{eq:objective}. This is the closest analogue to conventional MCS scheduling with a given CEV demand profile.

2) \emph{Baseline~2} fixes only the CEV work schedule $u_{i,e,t,a}$ to the reference schedule, while leaving the CEV charging intervals and all MCS decisions free by minimizing objective~\eqref{eq:objective}. Thus, comparing B2 with B1 highlights the value of optimizing CEV charging decisions, while comparing the proposed framework with B2 highlights the critical value of jointly optimizing the CEV work schedule.

The comparison is both fair and informative for two reasons. First, the reference schedules are feasible by construction. 
Second, all methods are evaluated under the same objective~\eqref{eq:objective}, so the performance differences highlight the value of jointly optimizing CEV charging and work schedules together with MCS operations.
\vspace{-1em}

\section{Results and Discussion}
\label{sec:results}
\subsection{Summary of Results Across All Scenarios} 
\label{subsec:all_scenarios}
Table~\ref{tab:cost_breakdown} reports the solver statistics and cost breakdown across the four scenarios, revealing four main trends. First, the proposed strategy achieves the lowest total cost in every scenario, reducing cost by $7$--$96\%$ relative to the best baseline, B2. Savings are smallest in Scenario~2 and largest in Scenario~3. The main driver is work completion: in Scenarios~1, 3, and~4, the fixed duty-cycle schedules leave $0.25$, $2$, and $0.5$~h of productive work unfinished because the rigid work-to-minimum/charge-to-full policy cannot be accommodated within the available working hours. In contrast, the proposed strategy completes all required work. The shortfall is greatest in Scenario~3, where one MCS serves two construction nodes and the fixed CEV schedule cannot adapt to limited availability. Thus, the baselines' lower MCS charging and energy costs reflect less completed work rather than greater efficiency.

\begin{figure*}[t]
\centering
\includegraphics[width=0.7\linewidth]{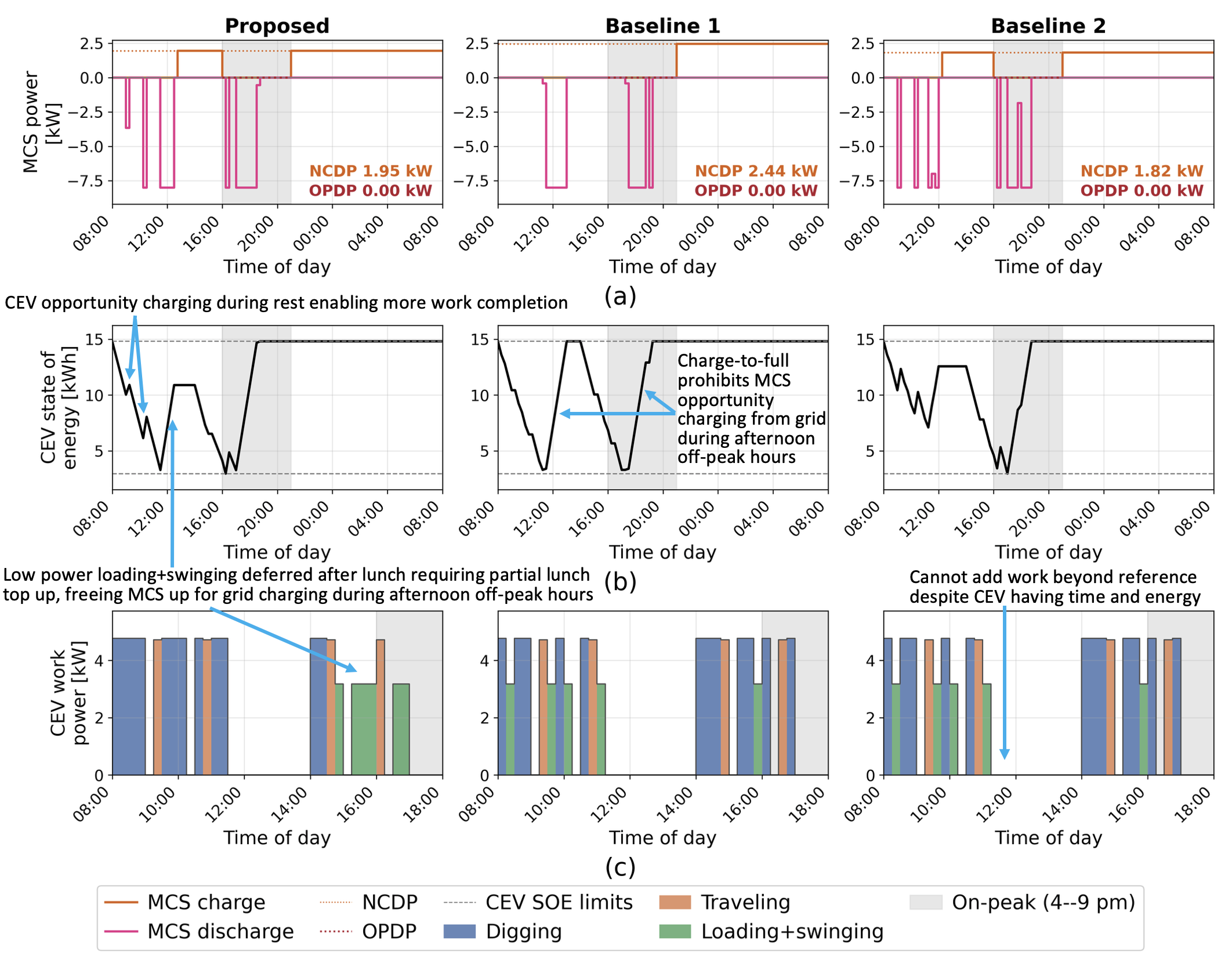}
\vspace{-1em}
\caption{Scenario~1 operation under the proposed strategy and the baselines B1 and B2 (columns): (a)~MCS charging (grid$\to$MCS, positive) and discharging (MCS$\to$CEV, negative) power; (b)~CEV SOE; and (c)~CEV work power, colored by subactivity; its time axis is zoomed to $08$:$00$--$18$:$00$~h for legibility. The $16$:$00$--$21$:$00$~h on-peak demand-charge window is shaded.}
\vspace{-1.5em}
\label{fig:scenario1_operation}
\end{figure*}

Second, joint optimization remains beneficial even when all work is completed. In Scenario~2, the proposed strategy reduces total cost by $7\%$ relative to B2 and by $14\%$ relative to B1. The improvement from B1 to B2 highlights the value of optimizing CEV charging time, which consistently reduces the NCDC. The remaining gap between B2 and the proposed strategy further highlights the benefit of jointly optimizing CEV work and charging schedules.

Third, on {tractability}, every baseline instance solves to proven optimality---within a second for B1, whose CEV work and charging schedules are both pinned, and within minutes for B2, since fixing the CEV-side binaries ($\mu$ and $u$) shrinks the branch-and-bound search space. The proposed formulation, which retains the full joint decision space, also solves to proven optimality in Scenarios~1--3 (in at most ${\sim}5.5$ minutes); and only in the largest case (Scenario~4) does it reach the one-hour limit with an $11.9\%$ MIP gap. As shown in Appendix~\ref{appendix_solver_non_optimality}, this gap is caused by a loose dual bound rather than a poor schedule: the best feasible objective stabilizes early, so the reported cost remains a reliable basis for comparison.

Finally, the proposed strategy sometimes yields the highest total CO$_2$ emissions, particularly in Scenario~3, because it supplies the additional energy required to complete work left unfinished by the baselines. When normalized by completed productive work, 
this disadvantage largely disappears and in some cases reverses: the proposed strategy has the lowest emissions per completed productive work-hour in Scenarios~1 and~4. In Scenario~2, where all methods complete the workload, the differences result solely from charging times relative to time-varying carbon intensity in Fig.~\ref{fig:electricity_carbon}. The proposed strategy outperforms B1, whose fixed intervals concentrate charging in carbon-intensive overnight hours, but not B2, which shifts more charging to solar-rich midday periods. In Scenario~3, the proposed strategy has the highest emissions per work-hour because completing the workload requires the full CEV travel prescribed by~\eqref{eq:travel_ub}--\eqref{eq:travel_lb}; this travel consumes energy without contributing productive work-hours, while the truncated baseline schedules omit part of both the travel and work.

\vspace{-1em}
\subsection{Interpreting Results of Scenario~1}
\label{subsec:scenario1_operation}

Table~\ref{tab:cost_breakdown} shows that for Scenario~1, most of the total cost gap between the proposed and baseline strategies is the \$$500$ missed-work penalty, which happens because of the rigid ``work-to-minimum/charge-to-full'' duty cycle of the reference.
 Figure~\ref{fig:scenario1_operation}(a) shows that all three strategies keep the MCS grid charging outside the $16$:$00$--$21$:$00$~h OP window, as OPDC are charged on top of NCDC. The OP hours are instead used for the CEV's end-of-day on-site recharge, and the MCS then spreads its required grid charging as flat blocks over off-peak hours. What separates them is the flexibility in off-peak time that the CEV schedule leaves for the MCS. 
 
 As observed in Fig.~\ref{fig:scenario1_operation}(b), in B1, the pinned CEV charging windows, i.e., a charge-to-full block over lunch ($11$:$15$--$13$:$00$~h) and an evening block from $17$:$00$--$19$:$15$~h hold the MCS at the construction node until late evening, so its only grid access is overnight, causing the highest NCDC. Its lower travel cost is because of avoiding an afternoon grid node trip (between $13$:$00$--$16$:$00$~h), which would save slightly less in demand charge than the extra travel costs. B2, which is free to re-time CEV charging, ends the lunch MCS-to-CEV top-up at $12$:$00$~h, releasing the MCS to capture an afternoon grid charging block ($12$:$15$--$16$:$00$~h) just before the OP window, which overall saves costs. The proposed strategy adopts this same charging template of frequent, short MCS--CEV top-ups with a pre-OP and an overnight flat grid block; its slightly higher NCDC reflects the extra ${\approx}1$~kWh it must deliver for the work the baselines leave unfinished.

The work completion and cost advantage of the proposed method comes from two sources. First, the proposed schedule converts the CEV's mandated rest breaks into charging intervals: with $15$-min top-ups at $09$:$00$~h and $10$:$15$~h (see Figs.~\ref{fig:scenario1_operation}(a),(b)), the CEV stays productive through the entire morning session. B1, which inherits both the CEV charging and work schedule from the rigid duty-cycle reference, instead forbids charging above the minimum SOE, so its rest breaks are idle and the CEV works ``to empty'' by $11$:$15$~h, and thereafter charging-to-full, stranding the remainder of the morning session, resulting in $0.25$~h of missed work it never recovers. B2 does retro-fit charging into those same rest breaks (see Figs.~\ref{fig:scenario1_operation}(a),(b)), which restores battery flexibility but cannot add work as its work schedule is fixed from the reference. 

Second, the proposed schedule sequences subactivities by power: it front-loads energy-intensive digging, performing $2.5$~h of its required $3$~h before lunch, and defers the lighter loading+swinging to the afternoon (see Fig.~\ref{fig:scenario1_operation}(c)). The lighter afternoon workload needs only a \emph{partial} lunch recharge, releasing the MCS to the grid in time for the pre-OP charging. The baseline schedules, having interspersed loading+swinging with digging from the morning onward, face a digging-heavy (energy-hungry) afternoon that 
still exhausts the battery by $17$:$00$~h with work unfinished. 
Even when operating costs excluding the missed-work penalty are normalized by completed productive work-hours, the proposed method (\$$15.2$/h) outperforms B1 (\$$16$/h) and B2 (\$$15.5$/h), demonstrating the benefit of power-aware CEV subactivity reordering. Scenario~2 is discussed in detail in Appendix~\ref{appendix_scenario_2}. 

\vspace{-0.5em}
\section{Conclusions and Future Work}
\label{sec:conclusions}
This paper presented a joint framework for scheduling CEV work and charging by MCSs. Using field data from a construction demonstration, we first identified the distinct subactivities of a compact electric excavator and developed and validated a constrained nonnegative least-squares model that recovers the average power consumption of each subactivity from coarse battery SOC telematics (within $17\%$ normalized mean absolute error); the accompanying dataset and code are released publicly. Based on these subactivity-resolved powers, we then formulated a mixed-integer program (MIP) that jointly optimizes CEV work schedules and MCS location, timing, and charging/discharging under economic and environmental objectives. Across realistic scenarios benchmarked against prevailing operational baselines, the proposed co-optimization consistently achieved the lowest cost ($7$--$96\%$ below the best-performing baseline), while also remaining computationally tractable. Future work will extend this formulation to a stochastic model predictive control framework to incorporate the uncertainty in CEV power draw, work requirements, electricity prices, and grid carbon intensity.



\vspace{-0.5em}
\enlargethispage{2\baselineskip}
\bibliographystyle{IEEEtran}
\bibliography{refs}

\newpage

\renewcommand{\theequation}{A.\arabic{equation}}
  \setcounter{equation}{0}  
  
\appendices

\section{Nomenclature}\label{appendix_nomenclature}
\begingroup
\renewcommand{\nomname}{}%
\printnomenclature
\endgroup

\section{CEV subactivity classification}\label{appendix_subactivity_classification}

\begin{table}[H]
\centering
\caption{Manually labeled subactivity taxonomy. $^{*}$ indicates subactivity not relevant for soil work.}
\label{tab:subactivity_taxonomy}
\begin{tabular}{@{}lp{5.6cm}@{}}
\toprule
Raw label & Description \\
\midrule
Digging      & Bucket excavation / ground breakout for trenching; surface leveling using back of bucket with arm/boom movement like trenching \\
Loading      & Filling / depositing material into the bucket \\
Swinging     & Rotation of the entire swinging body between loading and dumping points and back along with adjusting the bucket for dumping; surface leveling using swinging motion using side of bucket \\
Traveling   & Undercarriage (track) motion using traveling belt \\
Idling       & Resting, whether powered on or off \\
Mixing*       & In-place agitation of working material \\
\bottomrule
\end{tabular}
\end{table}

\section{SOC Quantization-Aware Observation Window}\label{appendix_windows}
\begin{figure}[H]
\centering
\includegraphics[width=\columnwidth]{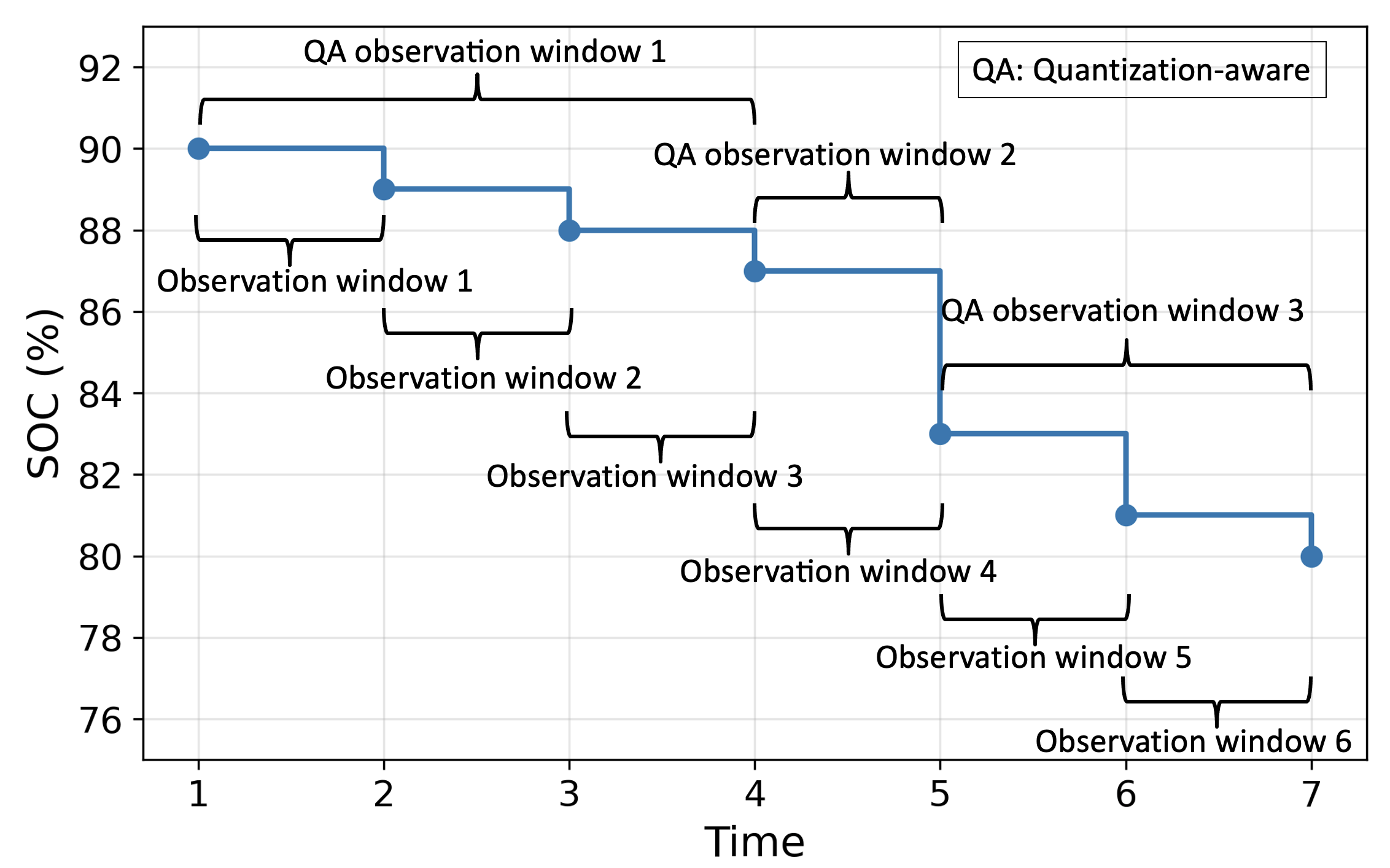}
\vspace{-1 em}
\caption{Observation window construction on measured battery SOC. Naive per-$1\%$ observation windows (lower brackets) place a boundary at every integer-percent SOC drop. Quantization-aware (QA) windows (upper brackets, $\tau=3\%$) instead open a new window only after the cumulative SOC has dropped by at least $\tau$, producing fewer but higher-signal-to-noise windows.}
\label{fig:window_quantized}
\end{figure}

\section{Subactivity Power Estimation for Decomposed Granite and Sand}\label{appendix_granite_sand}
The main paper focuses on soil, which dominates our dataset. For completeness, we report here the subactivity power estimation for the two remaining working materials, decomposed granite and sand, obtained with the identical pipeline: the SOC-quantization-aware window construction of Section~\ref{subsubsec:window} at $\tau=3\%$, the constrained non-negative least-squares estimator~\eqref{eq:nnls}, and the repeated $5$-fold cross-validation of Section~\ref{subsec:power_validation}.

Unlike soil and decomposed granite, work with sand exhibits the additional subactivity of mixing in Table~\ref{tab:subactivity_taxonomy}. To keep these small per-material datasets identifiable, we retain the Loading$+$Swinging clubbing of Section~\ref{subsubsec:collin}. The resulting full-data subactivity durations and cross-validated mean powers (with $95\%$ confidence intervals) are reported in Tables~\ref{tab:coeffs_granite} and~\ref{tab:coeffs_sand}, and the held-out prediction errors in Table~\ref{tab:metrics_granitesand}.

\begin{table}[H]
\centering
\scriptsize
\setlength{\tabcolsep}{2pt}
\renewcommand{\arraystretch}{1}
\caption{Estimated subactivity power for decomposed granite with $\tau=3\%$.}
\vspace{-0.8em}
\label{tab:coeffs_granite}
\begin{tabularx}{\columnwidth}{@{}lYYYY@{}}
\toprule
& Digging
& \makecell{Loading\\$+$Swinging}
& Traveling
& Idling \\
\midrule
\makecell[l]{Data share [\%]}
& 5
& 19
& 9
& 67 \\

\makecell[l]{Mean$\pm$std [kW]}
& $3.19\pm1.42$
& $3.34\pm0.52$
& $7.26\pm0.91$
& $0.00\pm0.00$ \\

\makecell[l]{95\% CI [kW]}
& $[0.64,\,6.38]$
& $[2.51,\,4.55]$
& $[5.73,\,9.16]$
& $[0.00,\,0.00]$ \\
\bottomrule
\end{tabularx}
\end{table}

\begin{table}[H]
\centering
\scriptsize
\setlength{\tabcolsep}{1pt}
\renewcommand{\arraystretch}{1}
\caption{Estimated subactivity power for sand with $\tau=3\%$.}
\vspace{-0.8em}
\label{tab:coeffs_sand}
\begin{tabularx}{\columnwidth}{@{}lYYYYY@{}}
\toprule
& Digging
& \makecell{Loading\\$+$Swinging}
& Traveling
& Mixing
& Idling \\
\midrule
\makecell[l]{Data share [\%]}
& 1
& 11
& 7
& 26
& 55 \\

\makecell[l]{Mean$\pm$std [kW]}
& $6.68\pm4.98$
& $4.57\pm0.87$
& $0.68\pm1.14$
& $5.01\pm0.21$
& $0.00\pm0.00$ \\

\makecell[l]{95\% CI [kW]}
& $[0.00,\,14.04]$
& $[3.46,\,6.64]$
& $[0.00,\,4.20]$
& $[4.69,\,5.59]$
& $[0.00,\,0.00]$ \\
\bottomrule
\end{tabularx}
\end{table}

\begin{table}[H]
\centering
\scriptsize
\setlength{\tabcolsep}{2pt}
\renewcommand{\arraystretch}{1}
\caption{Held-out prediction error with $\tau=3\%$ for decomposed granite and sand.}
\vspace{-0.8em}
\label{tab:metrics_granitesand}
\begin{tabularx}{\columnwidth}{@{}lYYYY@{}}
\toprule
& MAE [kWh]
& RMSE [kWh]
& MAPE [\%]
& \makecell{Normalized\\MAE [\%]} \\
\midrule
\makecell[l]{Decomposed granite}
& $0.152\pm0.054$
& $0.163\pm0.053$
& $25.2\pm9.3$
& $24.8\pm8.9$ \\

\makecell[l]{Sand}
& $0.125\pm0.050$
& $0.150\pm0.057$
& $19.9\pm8.6$
& $18.7\pm8.0$ \\
\bottomrule
\end{tabularx}
\end{table}

 These per-material fits are less accurate than the soil results of Section~\ref{subsec:power_validation}, for three related reasons. First, the datasets are much smaller: the construction yields only $11$ and $14$ energy-balance equations for decomposed granite and sand, respectively, against $25$ for soil, so each per-material system is far closer to being under-determined. Second, the soil workload is spread comparatively evenly across its estimated subactivities (see Table~\ref{tab:coeffs_soil}), giving every regressor substantial support, whereas the decomposed granite, and sand workloads are each dominated by a single subactivity---loading$+$swinging (see Table~\ref{tab:coeffs_granite}) for decomposed granite, and mixing (see Table~\ref{tab:coeffs_sand}) for sand, leaving the minority subactivities with little independent variation in the design matrix. Third, and as a direct consequence, these minority regressors are weakly identified for both decomposed granite and sand, i.e., their cross-validated intervals are wide. Thus, the decomposed granite and sand estimates are indicative, pending larger per-material datasets. 

\section{Case study settings}\label{appendix_case_study_settings}

\begin{figure*}[t]
\centering
\includegraphics[width=0.8\linewidth]{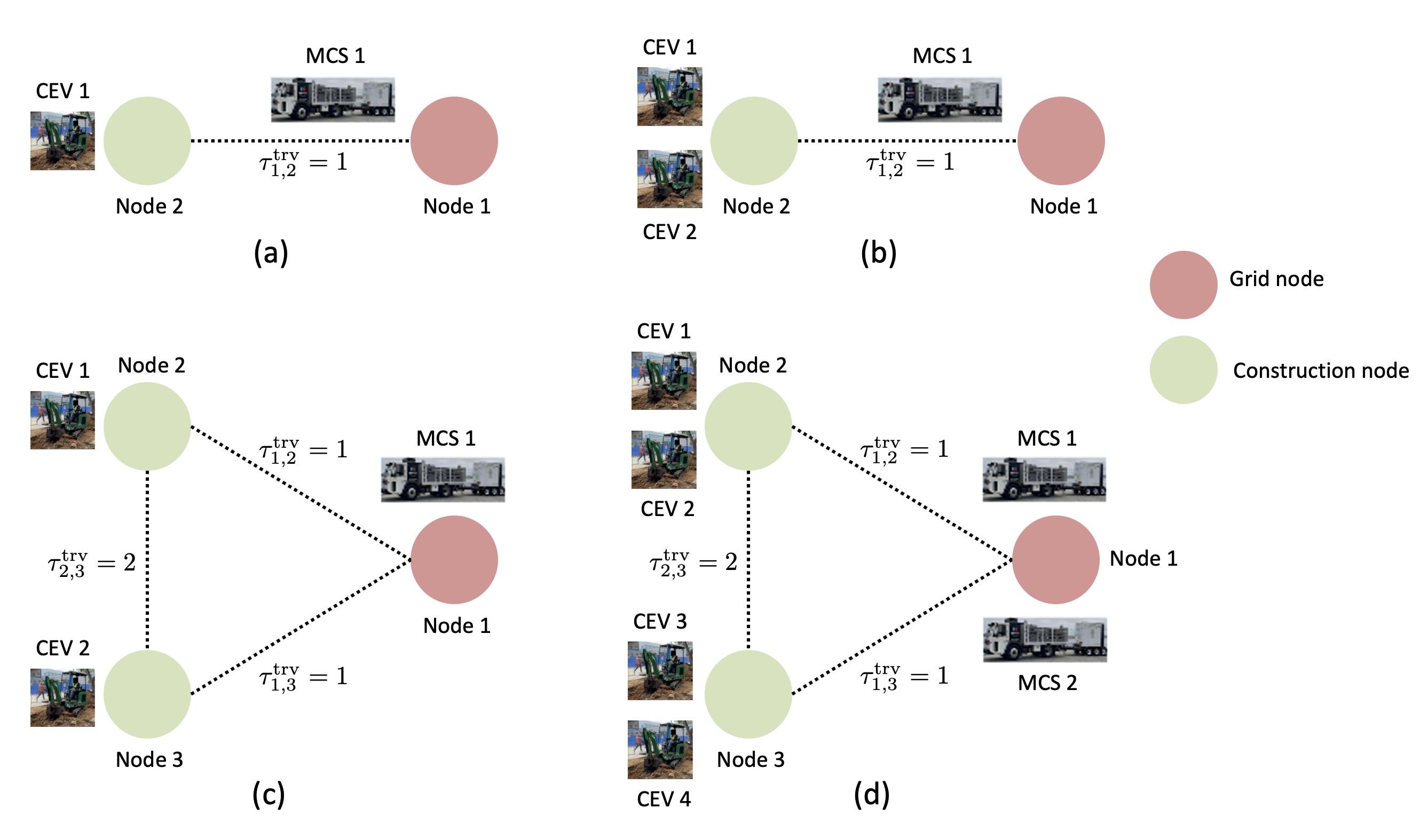}
\vspace{-0.5em}
\caption{Network topologies of the four case-study scenarios: (a)~Scenario~1, a single construction node with one CEV served by one MCS; (b)~Scenario~2, a single construction node with two CEVs and one MCS; (c)~Scenario~3, two construction nodes with one CEV each and one MCS; and (d)~Scenario~4, two construction nodes with two CEVs each and two MCSs. The arc labels give the inter-node travel times $\tau_{i,j}^{\mathrm{trv}}$ in number of time intervals.}
\vspace{-1em}
\label{fig:topology}
\end{figure*}

Figure~\ref{fig:topology} shows the network topologies of the four scenarios, and Table~\ref{tab:specifications} lists the CEV and MCS design specifications along with the operational parameters common to all of them.

\begin{table}[H]
\centering
\footnotesize
\setlength{\tabcolsep}{4pt}
\caption{Design specifications of the MCS and CEV, and the common scheduling-horizon and operational parameters used in the case study.}
\vspace{-1em}
\label{tab:specifications}
\resizebox{\columnwidth}{!}{%
\begin{tabular}{@{}llrl@{}}
\toprule
Symbol & Description & Value & Unit \\
\midrule
\multicolumn{4}{@{}l}{\textit{Mobile charging station (CleanGen~J250)}}\\
$\mathrm{SOE}_{m}^{\mathrm{MCS,ini}}$ & Initial \& terminal SOE        & $250$    & kWh   \\
$\mathrm{SOE}_{m}^{\mathrm{MCS,max}}$ & Maximum SOE                    & $250$    & kWh   \\
$\mathrm{SOE}_{m}^{\mathrm{MCS,min}}$ & Minimum SOE                    & $50$     & kWh   \\
$\text{CH}_{m}^{\mathrm{MCS}}$               & Grid charging capacity         & $31.25$  & kW    \\
$\text{DCH}_{m}^{\mathrm{MCS}}$              & Discharging capacity           & $80$     & kW    \\
$\text{DCH}_{m}^{\mathrm{plug}}$             & Per-plug discharging limit     & $52$     & kW    \\
$C_{m}^{\mathrm{plug}}$               & Number of outlet plugs         & $2$      & --    \\
$\eta_{m}$                            & Charging/discharging efficiency    & $0.95$   & --    \\
\addlinespace
\multicolumn{4}{@{}l}{\textit{Construction electric vehicle (JCB~19C-1E)}}\\
$C_{\mathrm{batt}}$                   & Usable battery capacity        & $14.8$   & kWh   \\
$\mathrm{SOE}_{e}^{\mathrm{CEV,ini}}$ & Initial \& terminal SOE        & $14.8$   & kWh   \\
$\mathrm{SOE}_{e}^{\mathrm{CEV,max}}$ & Maximum SOE                    & $14.8$   & kWh   \\
$\mathrm{SOE}_{e}^{\mathrm{CEV,min}}$ & Minimum SOE                    & $2.96$   & kWh   \\
$\text{CH}_{e}^{\mathrm{CEV}}$               & Charging acceptance rate       & $8$      & kW    \\
$\eta_{e}$                            & Charging efficiency    & $0.95$   & --    \\
\addlinespace
\multicolumn{4}{@{}l}{\textit{Scheduling \& operational parameters}}\\
$\Delta T$                            & Scheduling time step           & $0.25$   & h     \\
$t_{\mathrm{end}}$                    & Number of time intervals       & $96$     & --    \\
--                                    & Scheduling horizon length      & $24$     & h     \\
$\kappa^{\mathrm{seq}}$               & CEV precedence ratio (Dig $\rightarrow$ L+S) & $2$      & --    \\
$\kappa^{\mathrm{wt}}$                & CEV productive work\,:\,travel interval ratio & $4$      & --    \\
$t_{\mathrm{limit}}$                  & CEV work intervals before mandatory rest          & $4$      & -- \\
$\rho^{\mathrm{miss}}$                & Missed-work penalty                           & $2000$   & \$/h  \\
$\rho^{\mathrm{travel}}$              & MCS towing labor cost                         & $20$    & \$/h  \\
$\lambda^{\mathrm{em}}$               & Nominal carbon-emissions price                        & $0.05$     & \$/kg\,CO$_2$ \\
$\tau_{i,j}^{\mathrm{trv}},\, i \in \mathcal{N}^g, j \in \mathcal{N}_c$              & Travel time intervals between grid and construction nodes                        & $1$     &  --  \\
$\tau_{i,j}^{\mathrm{trv}},\, i,j \in \mathcal{N}^c, i\neq j$              & Travel time intervals between construction nodes                        & $2$     &  --  \\
\addlinespace
\multicolumn{4}{@{}l}{{\textit{MCS initial position} (parked at or departing from a grid node at $t{=}1$):}}\\
\multicolumn{4}{@{}l}{{\quad $\sum_{i\in\mathcal{N}^{g}} z_{m,i,1}+\sum_{i\in\mathcal{N}^{g}}\sum_{\substack{j\in\mathcal{N},\, j\neq i}}x_{m,i,j,1}=1, \;\;\forall m\in\mathcal{M}$}}\\
\bottomrule
\end{tabular}%
}
\end{table}

\section{Solver Convergence and Non-Optimality}
\label{appendix_solver_non_optimality}

\begin{figure}[h]
\centering
\includegraphics[width=\columnwidth]{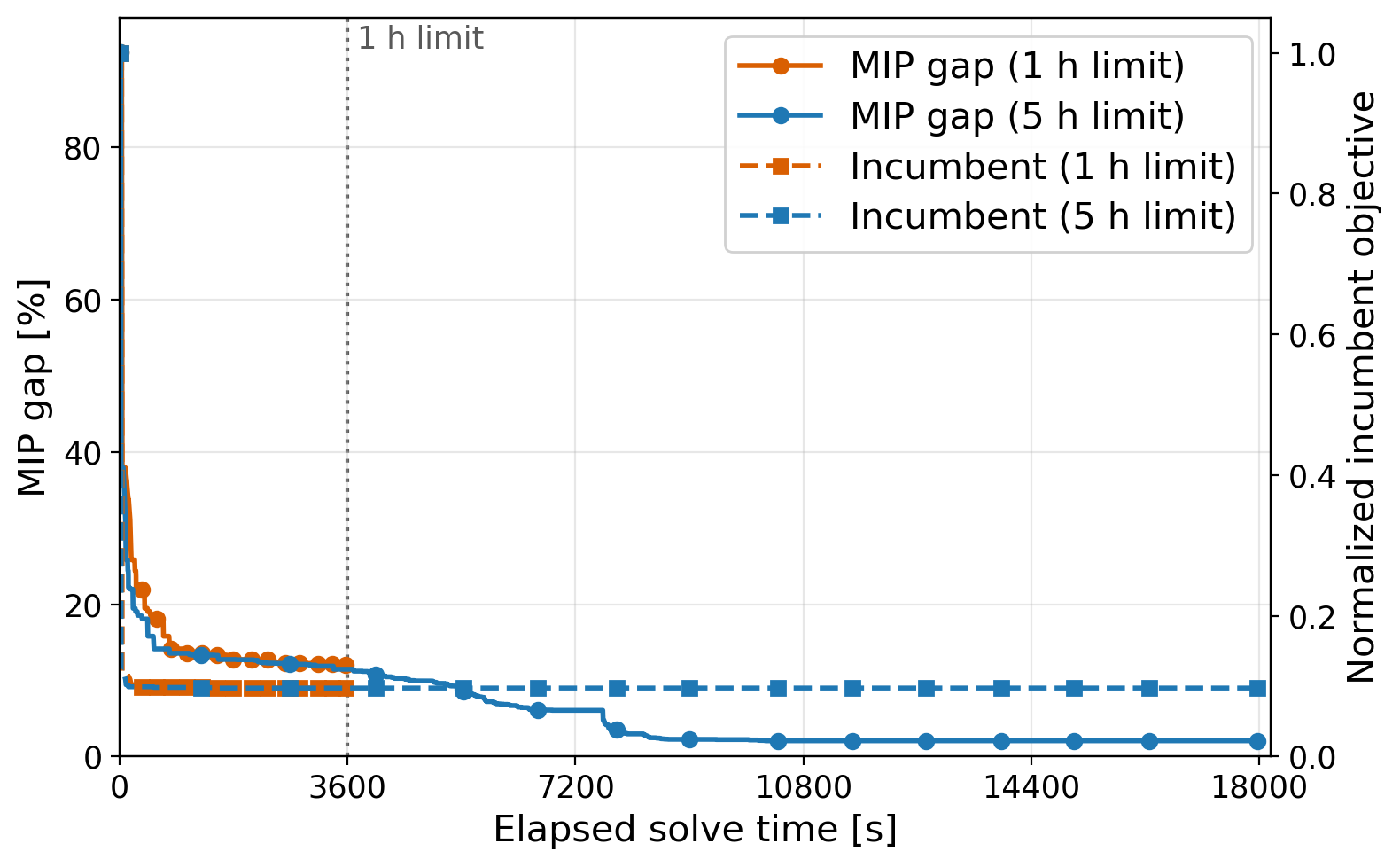}
\vspace{-2em}
\caption{Solver convergence of the proposed strategy in Scenario~4 under the one-hour and five-hour time limits: MIP gap and normalized incumbent objective, i.e., the best feasible objective value at each instant, normalized by its largest value over the run, versus elapsed solve time. The dotted vertical line marks the one-hour limit.}
\label{fig:mip_convergence}
\end{figure}

Figure~\ref{fig:mip_convergence} shows the solver convergence for the only instance in Table~\ref{tab:cost_breakdown} that is not solved to proven optimality within the one-hour limit: the proposed strategy in Scenario~4. The figure establishes that the residual $11.9\%$ MIP gap is an artifact of a loose \emph{dual} bound rather than of a poor \emph{incumbent} (i.e., the best feasible solution found). In the one-hour run, the normalized incumbent objective, i.e., the best feasible cost found at each instant normalized by its largest value over the run, collapses within the first few hundred seconds and never improves thereafter. This implies that the schedule reported in Table~\ref{tab:cost_breakdown} is found early, and the gap's slow decline after ${\sim}800$~s is driven entirely by the dual bound inching upward, not by better incumbents. Re-running the instance with a five-hour limit corroborates this: the solver settles on the \emph{same} incumbent, and hence the same cost as reported in Table~\ref{tab:cost_breakdown}, with its last improvement at ${\sim}26$~minutes, while the MIP gap shrinks from $12.8\%$ to $2.1\%$ purely through further dual-bound tightening. The one-hour limit therefore does not bias the reported cost.

The looseness of the dual bound in Scenario~4 is due to~\emph{symmetry}. Scenario~4 comprises two identical MCSs and two identical CEVs at each of the two construction nodes, so any incumbent schedule admits equivalent permutations (swapping the MCSs' routes, or the CEV indices within a site) with identical cost. Branch-and-bound must implicitly enumerate these symmetric subtrees, which slows the tightening of the dual bound without there being any better incumbent to find. The reported cost is therefore effectively converged, which makes the cost comparisons in Table~\ref{tab:cost_breakdown} valid.

\section{Interpreting results of Scenario~2}
\label{appendix_scenario_2}

\begin{figure*}[t]
\centering
\includegraphics[width=0.7\linewidth]{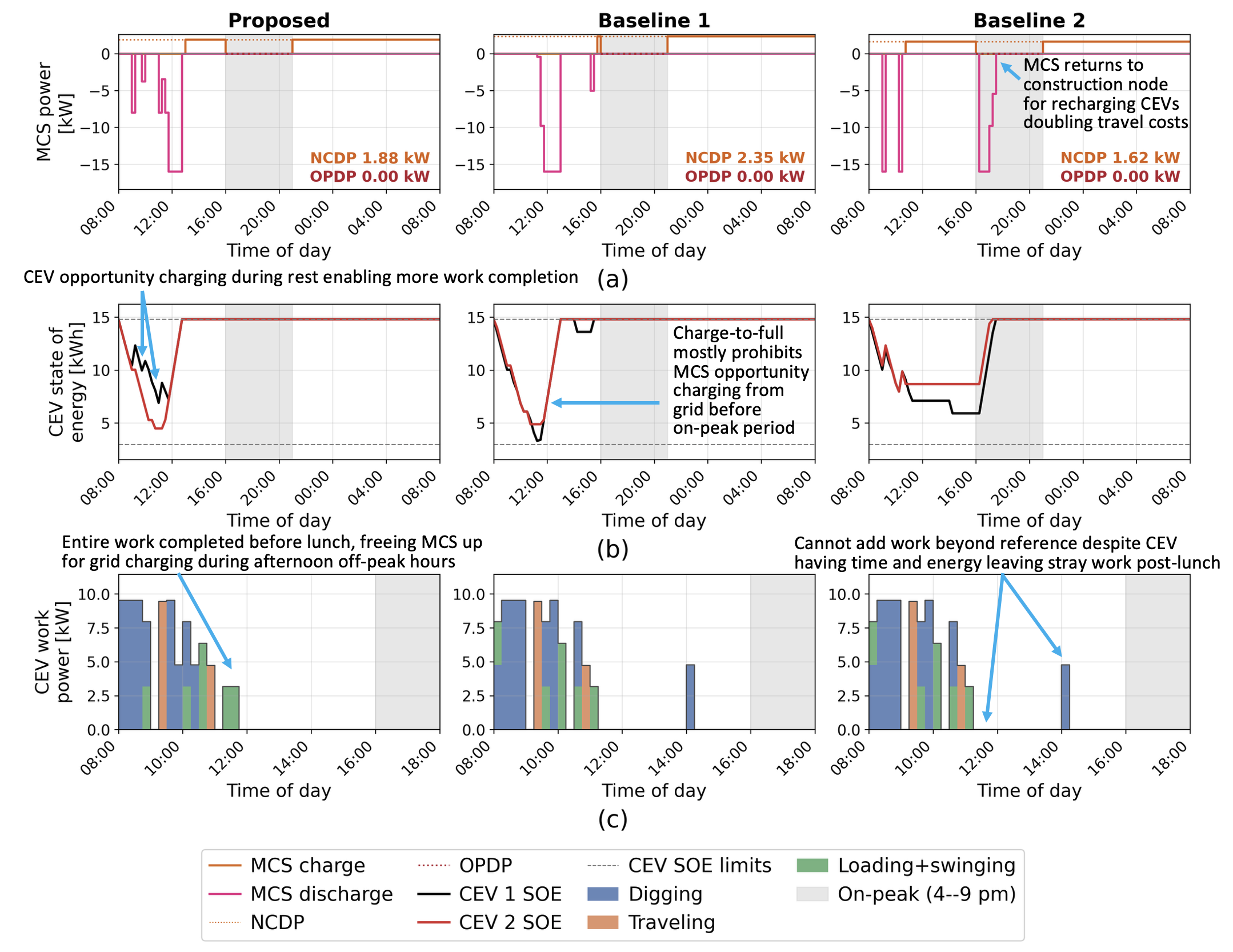}
\vspace{-1em}
\caption{Scenario~2 operation under the proposed strategy and the baselines B1 and B2 (columns): (a)~MCS charging (grid$\to$MCS, positive) and discharging (MCS$\to$CEV, negative) power; (b)~SOE of the two CEVs; and (c)~site-total CEV work power (stacked over the two CEVs), colored by subactivity; its time axis is zoomed to $08$:$00$--$18$:$00$~h for
legibility. The $16$:$00$--$21$:$00$~h on-peak demand-charge window is shaded.}
\vspace{-1em}
\label{fig:scenario2_operation}
\end{figure*}

Scenario~2 is the only case in which the reference CEV schedule completes the full workload, making it the cleanest test of the joint optimization's value beyond work completion: the proposed strategy still attains the lowest total cost (Table~\ref{tab:cost_breakdown}). 

Figure~\ref{fig:scenario2_operation}(a) shows that all three strategies keep the MCS grid charging out of the $16$:$00$--$21$:$00$~h OP window and spread it as flat off-peak blocks. The decisive difference between the proposed method and the baselines is the reference schedule's single residual work interval at $14$:$00$--$14$:$15$~h (see Fig.~\ref{fig:scenario2_operation}(c)), which is an artifact of the rigid reference duty cycle. This afternoon work combined with the requirement that the CEVs end the day fully charged, forces the CEVs' final top-up to occur \emph{after} $14$:$15$~h under both B1's and B2's MCS dispatch. B1, whose charging windows are additionally pinned, holds the MCS at the construction node until later afternoon, confining its grid access essentially to overnight and forcing the highest flat draw and hence the highest NCDC. B2, free to re-time CEV charging, shifts the top-ups into the rest breaks and into the OP window (when grid charging is priced out), freeing the MCS for a long pre-OP grid block ($10$:$45$--$16$:$00$~h) that yields the lowest flat draw and NCDC. However, the pinned afternoon work interval forces it to return to the site to recharge the CEVs post work, incurring an extra round trip and doubling its travel cost.

The proposed strategy dissolves this dilemma by re-ordering the work itself: with the two CEVs working in parallel and topped up during their rest breaks, the entire $4.5$~h workload completes within the morning session and both CEVs are charged to full by $12$:$45$~h (see Figs.~\ref{fig:scenario2_operation}(b),(c)). The MCS then departs for the grid once and never returns, charging in pre-OP hours and overnight with the least travel cost and the second-lowest flat draw, leading to least overall costs. 


\end{document}